\documentclass[10pt,aps,prd,reprint,nofootinbib,floats,floatfix,amsfonts,amssymb,amsmath,preprintnumbers,notitlepage,superscriptaddress,longbibliography]{revtex4-1}
\usepackage{graphicx}
\usepackage[utf8]{inputenc}
\usepackage{aas_macros}
\usepackage[british]{babel}
\usepackage{isomath}
\usepackage[protrusion=true,tracking=true,kerning=true,spacing=true,final,babel=true]{microtype}
\usepackage{physics}
\usepackage{xcolor}
\def\TimeToDetectionStrong{\ensuremath{2.1^{+1.7}_{-0.7}}~years}
\def\TimeToDetectionStrongNoisy{\ensuremath{2.1^{+1.2}_{-0.9}}~years}
\def\TimeToDetectionNoSchumann{\ensuremath{1.5^{+0.9}_{-0.6}}~years}

\usepackage{url}
\usepackage{nth}
\newcommand{\bs}[1]{\boldsymbol{#1}}
\newcommand{\logodds}[2]{\ln\mathcal{B}_{\rm #2}^{\rm #1}}
\newcommand{\odds}[2]{\mathcal{B}_{\rm #2}^{\rm #1}}

\begin{document}

\begin{flushleft}
KCL-PH-TH/2020-37
\end{flushleft}

\title{Detecting a stochastic gravitational-wave background in the presence of correlated magnetic noise}
\author{Patrick~M.~Meyers}
\email{pat.meyers@unimelb.edu.au}
\affiliation{School of Physics, University of Melbourne, Parkville, VIC 3010, Australia}
\affiliation{OzGrav, University of Melbourne, Parkville, VIC 3010, Australia}
\author{Katarina~Martinovic}
\email{katarina.martinovic@kcl.ac.uk}
\affiliation{Theoretical Particle Physics and Cosmology Group, \, Physics \, Department, \\ King's College London, \, University \, of London, \, Strand, \, London \, WC2R \, 2LS, \, UK}

\author{Nelson Christensen}
\email{nelson.christensen@oca.eu}
\affiliation{Artemis, Universit\'e C\^ote d'Azur, Observatoire de la C\^ote d'Azur, CNRS, Nice 06300, France}
\author{Mairi~Sakellariadou}
\email{mairi.sakellariadou@kcl.ac.uk}
\affiliation{Theoretical Particle Physics and Cosmology Group, \, Physics \, Department, \\ King's College London, \, University \, of London, \, Strand, \, London \, WC2R \, 2LS, \, UK}
\date{\today}

\begin{abstract}
A detection of the stochastic gravitational-wave background (SGWB) from unresolved compact binary coalescences could be made by Advanced LIGO and Advanced Virgo at their design sensitivities. However, it is possible for magnetic noise that is correlated between spatially separated ground-based detectors to mimic a SGWB signal. In this paper we propose a new method for detecting correlated magnetic noise and separating it from a true SGWB signal. A commonly discussed method for addressing correlated magnetic noise is coherent subtraction in the raw data using Wiener filtering.  The method proposed here uses a parameterized model of the magnetometer-to-strain coupling functions, along with measurements from local magnetometers, to estimate the contribution of correlated noise to the traditional SGWB detection statistic. We then use Bayesian model selection to distinguish between models that include correlated magnetic noise and those with a SGWB. Realistic simulations are used to show that this method prevents a false SGWB detection due to correlated magnetic noise. We also demonstrate that it can be used for a detection of a SGWB in the presence of strong correlated magnetic noise, albeit with reduced significance compared to the case with no correlated noise. Finally, we discuss the advantages of using a global three-detector network for both identifying and characterizing correlated magnetic noise.
\end{abstract}{}
\maketitle
\section{Introduction}
\label{sec:intro}

During their first two observational runs, Advanced LIGO~\cite{TheLIGOScientific:2014jea} and Advanced Virgo~\cite{TheVirgo:2014hva} detected gravitational wave (GW) signals from 10 binary black hole mergers, and one binary neutron star merger~\cite{LIGOScientific:2018mvr}. During the third observation run (O3), numerous low-latency alerts for binary black hole, binary neutron star, and neutron star-black hole mergers have been sent out to astronomers~\cite{gracedb_public}. Exceptional events from O3 are now being published, including new compact binary mergers that could be neutron star-black hole mergers~\cite{Abbott:2020uma,Abbott_2020_190814}. These detections have already made a broad-reaching impact on stellar astrophysics, the study of dense nuclear matter, and beyond. In the coming years, one of the main targets of ground-based interferometeric GW detectors will be a detection of the stochastic gravitational-wave background (SGWB). In this paper, we address a potential hurdle faced on the way to the eventual detection of such a background. We expect a SGWB from unresolved compact binary coalescences (CBCs) could be detectable by the time Advanced LIGO and Advanced Virgo reach design sensitivity~\cite{Abbott:2017xzg}. Other sources, both astrophysical and cosmological, can contribute to the SGWB, the most of exciting of which include GWs from the early Universe~\cite{Christensen:2018iqi}. SGWB searches can also complement transient GW searches,
through, e.g. searching for alternative polarisations of GWs~\cite{Abbott:2017oio,Callister:2017ocg,Abbott:2018utx}, and can be used together with transient detections to constrain the star formation history of the Universe~\cite{Callister:2020arv}.

Unlike for transient signals, searches for a SGWB require long integration times because the signal is much smaller than the intrinsic detector noise. We search for the SGWB by cross-correlating outputs from two or more widely-separated detectors, and when there are no correlated noise sources between the detectors, the only limiting factor of the search is total observation time \cite{PhysRevD.46.5250,Romano:2016dpx}. In the presence of noise that is correlated between the detectors, however, we must accurately estimate and separate the relative strengths of the correlated noise and the SGWB.

Correlated noise in GW detectors caused by the Earth's electromagnetic field, in the form of the Schumann resonances, could be comparable to the sensitivity of SGWB searches performed by the advanced detector network in the near future~\cite{PhysRevD.46.5250,Thrane:2013npa, Thrane:2014yza}. Analytic models of the impact of correlated noise in GW detectors have been explored in recent work, such as~\cite{Himemoto:2017oiw, Himemoto2019}. Meanwhile, the focus of most attempts at mitigating the effects of correlated detector noise on SGWB searches has centered on Wiener filtering of the correlated signal using local environmental sensors~\cite{Thrane:2013npa,Thrane:2014yza,Coughlin:2016vor,Cirone:2018guh}. In addition, a method for validating a potential SGWB signal once it is detected using geodesy~\cite{Callister:2018ogx} has also been proposed. The geodesy method provides a check on whether a proposed SGWB signal is consistent with an isotropic SGWB or if it is more consistent with environmental disturbances. Such a method offers complementary information to approaches that attempt to subtract or mitigate correlated noise.

In this paper, we take a different tact. We model the contribution of correlated magnetic noise from Schumann resonances to the frequency-domain SGWB estimator used by most searches~\cite{allen:1999ied}. We propose a method to simultaneously detect correlated magnetic noise and a SGWB using local on-site magnetometers and current SGWB search data products. We then demonstrate this method using realistic time-domain and frequency-domain synthetic data sets with varying levels of correlated magnetic noise. Such a method offers an alternative to Wiener filering, but could also be used on data that has already had Wiener filtering subtraction applied, given that Wiener filtering in the low signal-to-noise regime can result in imperfect subtraction~\cite{Cirone:2018guh}.

The rest of this paper is organized as follows. In Section~\ref{sec:sgwb_searches}, we introduce the cross-correlation statistic used in SGWB searches, and highlight complications introduced by correlated detector noise. In Section~\ref{sec:simulating_data_with_correlated_noise}, we explain the Schumann resonances and their coupling to the detectors, and we present the way we model this coupling. We then present a method of simulating synthetic time series data that includes a correlated magnetic spectrum in a multi-detector network. In Section~\ref{sec:pe_ms}, we discuss a model for the SGWB search statistic that includes correlated magnetic noise, and demonstrate how we use that model to co-detect the presence of correlated magnetic noise and a SGWB. We present results on synthetic data in Section~\ref{sec:results}, and finish with a brief discussion and suggestions for future work in Section~\ref{sec:discussion}.

\section{SGWB and Search Methods}
\label{sec:sgwb_searches}
If we assume the SGWB is isotropic, Gaussian, stationary, and unpolarized, then it is fully characterized by the dimensionless energy density per logarithmic frequency interval
\begin{equation}
   \Omega_{\rm gw}(f) = \frac{1}{\rho_c} \frac{\text{d} \rho_{\rm gw}(f)}{\text{d}  \text{ln}(f)},
\end{equation}
where $\textrm{d}\rho_{\rm gw}$ is the GW energy density in the frequency interval $\ln f$ to $\ln f+\mathrm{d} \ln f$, and $\rho_c=3H_0^2 c^2/(8\pi G)$ is the critical energy density to close the Universe. It is common to model the SGWB spectrum as a power law:
\begin{equation}
\label{eq:omega_gw_power_law}
   \Omega_{\rm gw}(f) = \Omega_\alpha \,\left(\frac{f}{f_{\rm ref}}\right)^{\alpha},
\end{equation}
where $\Omega_{\alpha}$ is the amplitude at a reference frequency, $f_{\rm ref}$, and $\alpha$ is the spectral index. We will use $f_{\rm ref} = 25 $ Hz.

Unresolved CBCs give a background spectrum with $\alpha=2/3$; slow roll inflation models and cosmic strings predict $\alpha=0$. It is also common to consider a model that is flat in GW power, which corresponds to $\alpha=3$, to mimic signals like those from phase transitions and supernovae~\cite{Christensen:2018iqi}.
Recent estimates suggest that the SGWB could be detected by the Advanced LIGO and Advanced Virgo detector network once these detectors reach design sensitivity and integrate for $\cal O(\rm years)$~\cite{Abbott:2017xzg}.

In what follows, we consider a SGWB search that uses a cross-correlation estimator that is optimal for a Gaussian, stationary, unpolarized and isotropic background. Our estimator, $\hat C_{ij}(f)$, for the SGWB measured from detectors $i$ and $j$ is
\begin{align}
\label{eq:sgwb_searches:c_hat_definition}
\hat C_{ij}(f;t) = \frac{2}{T}\frac{\textrm{Re}[\tilde s_i^*(f;t)\tilde s_j(f;t)]}{\Gamma_{ij}(f)S_0(f)},
\end{align}
where $\tilde s_i(f;t)$ is the Fourier transform of the strain time series in detector $i$ starting at time $t$, $\Gamma_{ij}(f)$ is the normalized overlap reduction function (ORF)~\cite{PhysRevD.46.5250,mingarelli_understanding_omega_gw} between detectors $i$ and $j$, $T$ is the duration over which the Fourier transform is taken, and $S_0(f)$ is the spectral shape for a SGWB that is flat in energy density, $S_0(f) = 3 H_0^2 /( 10\pi^2 f^3)$.

In the limit where the total GW strain amplitude in detector $i$, $\tilde h_i(f)$, is much less than the intrinsic detector noise, $\tilde n_i(f)$, the variance of $\hat C_{ij}(f;t)$ is given by
\begin{align}
\label{eq:sgwb_searches:sigma_on_c_hat}
\sigma_{ij}^2(f;t) = \frac{1}{2\Delta f T}\frac{P_i(f;t) P_j(f;t)}{\Gamma_{ij}(f)^2S_0(f)^2},
\end{align}
where $P_i(f;t)$ is the one-sided power spectral density (PSD) of detector $i$ between times $t$ and $t+T$, and $\Delta f$ is the frequency resolution.

In general, Eq.~(\ref{eq:sgwb_searches:c_hat_definition}) and Eq.~(\ref{eq:sgwb_searches:sigma_on_c_hat}) are estimated for many short time-segments of $T = 192$~s and these segments are optimally combined in a post-processing step given by
\begin{align}
   \hat C_{ij}(f) &= \frac{\sum_k \hat C_{ij, k}(f) \sigma_{ij,k}^{-2}(f)}{\sum_k \sigma_{ij,k}^{-2}(f)},\\
   \sigma_{ij}(f) &= \left(\sum_k \sigma_{ij,k}^{-2}(f)\right)^{-1/2},
\end{align}
where $k$ indexes the time segments. For a set of $N_t$ time segments starting at times $\{t_{k}\}_{k=1}^{k=N_t}$, we have defined $\hat C_{ij, k}(f) = \hat C_{ij}(f; t_k)$, and likewise for its variance.

It is worth considering the expectation value of the estimator, $\langle \hat C_{ij}(f)\rangle$, in some detail (we will suppress the time-dependence for brevity).
Let us assume that $\tilde{s}_i(f)$ can be written as
\begin{align}
\tilde{s}_i(f)=\tilde{h}_i(f) + \tilde{n}_i(f) \color{black},
\end{align}
where $\tilde n_i(f)$ is the Fourier transform of the instrument noise in detector $i$, and
\begin{align}
\tilde h_i(f) = \sum_A\int \textrm{d}^2\hat r\, F_i^A(f, \hat r)\tilde h_A(f, \hat r)e^{-2\pi i f \vec x_i \cdot \vec r/ c}
\end{align}
is the total GW signal in detector $i$ located at $\vec x_i$. Here $F_i^A(f, \hat r)$ is the response of detector $i$ to a plane-wave traveling in direction $\hat r$ with polarization $A$, and $\tilde h_A(f, \hat r)$ is the Fourier amplitude of that plane wave.
Consequently,
\begin{eqnarray}
\label{eq:sgwb_searches:si_sj_expansion}
  \langle \tilde{s}_i^*(f)\tilde{s}_j(f') \rangle &=& \langle \tilde{h}_i^*(f)\tilde{h}_j(f') \rangle +  \langle \tilde{h}_i^*(f)\tilde{n}_j(f') \rangle\nonumber\\
    &&+ \langle \tilde{n}_i^*(f)\tilde{h}_j(f') \rangle + \langle \tilde{n}_i^*(f)\tilde{n}_j(f') \rangle.
\end{eqnarray}
If we assume that the SGWB is isotropic, Gaussian, stationary and unpolarized, then it is well-described by a single power spectral density $S_{\rm gw}(f)$,
\begin{equation}
\label{eq:sgwb_searches:hi_hj_assumption}
  \langle \tilde{h}_i^*(f)\tilde{h}_j(f') \rangle \, =  \frac{1}{2}\delta_T(f-f')\Gamma_{ij}(f) S_{\rm gw}(f),
\end{equation}
where $\delta_T(f-f')$ is the finite-time approximation to the dirac delta function, and $S_{\rm gw}(f)$ is related to the dimensionless energy density as follows
\begin{equation}
\label{eq:sgwb_searches:s_gw_definition}
S_{\rm gw}(f) = \frac{3 H_0^2}{10 \pi^2} \, \frac{\Omega_{\rm gw}(f)}{f^3}.
\end{equation}
Note that, for the existing detectors, the overlap reduction function, $\Gamma_{ij}(f)$, accounts for all the geometric factors that come into play when cross-correlating data from different detectors ~\cite{PhysRevD.46.5250}.

Combining Eqs.~(\ref{eq:sgwb_searches:si_sj_expansion})--(\ref{eq:sgwb_searches:s_gw_definition}), substituting into Eq.~(\ref{eq:sgwb_searches:c_hat_definition}), and then including the time-dependence again, we find
\begin{align}
\label{eq:sgwb_searches:c_hat_expectation_value}
\langle \hat C_{ij}(f;t)\rangle = \Omega_{\rm gw}(f) + 2\,\textrm{Re}\left[\frac{\langle \tilde n_i^*(f;t)\tilde n_j(f;t)\rangle}{T\Gamma_{ij}(f)S_0(f)}\right],
\end{align}
where we have assumed that the GW signal and the intrinsic noise are uncorrelated, $\langle \tilde h_i^*(f)\tilde n_j(f')\rangle = 0$, and that the noise in each frequency bin is independent. It is clear from (\ref{eq:sgwb_searches:c_hat_expectation_value}) that in the absence of correlated noise, i.e. $\langle \tilde n_i^*(f)\tilde n_j(f)\rangle=0$, $\langle \hat C_{ij}(f)\rangle$ is an estimator for $\Omega_{\rm gw}(f)$. However, this is not the case when $\langle \tilde n_i^*(f)\tilde n_j(f)\rangle \neq 0$.

Schumann resonances are a potential source of correlated magnetic noise. An estimate of the correlated magnetic noise contribution in the isotropic SGWB search using data from Advanced LIGO's first and second observing runs indicates that it is not yet an issue for current searches~\cite{Abbott:2019owe}. However, as detectors grow more sensitive, this will likely change, and the magnetic noise budget could dominate the signal \cite{Thrane:2014yza}. Hence, a careful treatment of correlated magnetic noise is of vital importance.

\section{Simulating GW data with correlated noise}
\label{sec:simulating_data_with_correlated_noise}
In this section, we discuss how we simulate GW data that is contaminated with correlated noise due to the Schumann resonances. In \ref{ssec:schumann} we discuss the Schumann resonances and their general properties. In \ref{subsec:Coupling} we present a model for the coupling of magnetic fields into GW detectors. In \ref{subsec:simulating_data} we show how to simulate multiple data streams that have correlated Gaussian noise components, and then we apply that method to our specific use case.

\subsection{Schumann Resonances}
\label{ssec:schumann}

\begin{figure}[h]
    \centering
    \includegraphics[width=.8\linewidth]{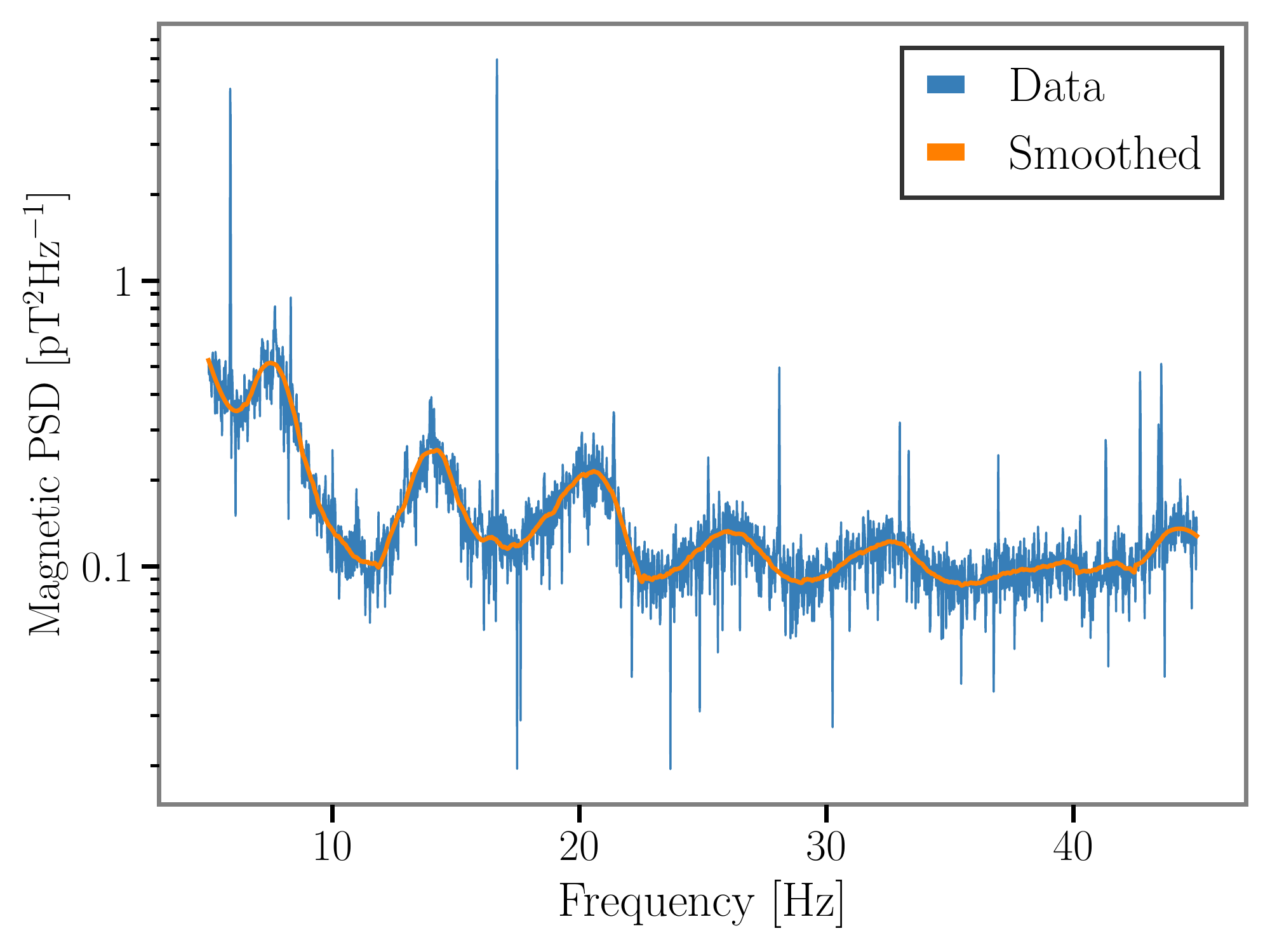}
    \caption{Power spectral density of magnetometer data near the Advanced Virgo detector. The blue is the inverse-averaged power spectral density for many 32~s chunks of data for the period from 00:00--02:00 UTC on July 9th, 2019. We use inverse averaging to account for possible magnetic transients that occur during this time. We produce the orange curve by removing the large, narrow spectral features and applying a smoothing filter. We can clearly see five harmonics of the Schumann resonances. The large, narrow spectral features are caused by local magnetic noise on site at Virgo. }
    \label{fig:SR}
\end{figure}

In 1952, Schumann predicted the existence of global extremely low frequency (ELF) peaks in the electromagnetic field of the Earth, which were subsequently observed~\cite{1952ZNatA...7..149S, 1954NW.....41..183S}. The resonances are eigenmodes of the conducting spherical cavity formed by the surface of the Earth and its ionosphere, and are excited by lightning discharges \cite{2016Atmos...7..116P}. The first harmonic, which corresponds to the circumference of the Earth, is at 7.8 Hz, and the subsequent harmonics are at 14 Hz, 20.8 Hz and 27.3 Hz. The first mode has the strongest resonance peak, with each consecutive peak being weaker than the previous one. In Figure~\ref{fig:SR}, we show the power spectral density seen in low-noise magnetometers on-site at the Advanced Virgo detector. We can clearly see the first five harmonics of the Schumann resonances. 
There is a diurnal variation in the amplitude of the Schumann resonances that corresponds to electrical storms that start at similar times and places each day~\cite{Sentman,PRICE20041179}.   The amplitude of the resonance peaks can vary by as much as a factor of two between the loudest and quietest times of the day, depending on the time of year and the location~\cite{PRICE20041179,ZHOU201386,2016Atmos...7..116P}. What is shown in Figure~\ref{fig:SR} represents a trough in the height of the peaks over the course of the day at Virgo. Despite this diurnal variation, we will model the spectrum as stationary in this paper for simplicity.

{}

The Schumann resonances, being global excitations, are coherent across the $\mathcal{O}(1000~\rm{km})$ distance scales between GW detectors~\cite{Coughlin:2016vor,Cirone:2018guh}. We model the time-series induced in magnetometers from the Schumann resonances as Gaussian, stationary,
and unpolarized, with a power-spectral density that can be described by a set of Lorentzians centered around the main harmonics. We assume that the data in two magnetometers, $\tilde m_i(f)$ and $\tilde m_j(f)$, have a cross-spectral density given by
\begin{align}
    \langle \tilde m_i^*(f) \tilde m_j(f')\rangle = \frac{1}{2}\delta_T(f-f')\gamma^M_{ij}(f)M(f),
    \label{eq:mag_coh}
\end{align}
where $M(f)$ is the correlated power spectral density and $\gamma^M_{ij}(f)$ is the magnetic analogue to the GW ORF, $\Gamma_{ij}$. This model is equivalent to Eq. (23) of~\cite{Himemoto:2017oiw}, and we refer the reader to that paper for an in-depth discussion of the model.

\subsection{Coupling to detectors}
\label{subsec:Coupling}

Magnetic fields can induce noise in GW detectors by coupling to metallic materials in the suspension system of the detector, or by inducing currents in the cabling.
The magnetic coupling is estimated by injecting magnetic noise into the detector, and measuring the detector's response, and the response of the witness magnetometers near the detectors. Peaks in the detectors' strain channels are related to the peaks in the magnetometer channels via the coupling function, $T(f)$ \cite{Thrane:2013npa}:

\begin{equation}
\label{eq:transfer}
\tilde{n}(f) = T(f) \tilde{m}(f).
\end{equation}
The exact frequency dependence of the coupling function is uncertain, and it can change over the course of a long observation run~\cite{ligo_logbook}. Throughout this paper, we will assume that the coupling is constant in time, is well-described by a power law, and is real. It takes the form
\begin{equation}
\label{eq:t(f)}
    T(f) = \kappa \, \left(\frac{f}{10 \textrm{ Hz}}\right)^{-\beta} \times 10^{-23} \; \text{strain/pT},
\end{equation}
where $\kappa$ is the amplitude of the coupling at $10$~Hz and $\beta$ is the spectral index of the power law.
In \cite{Thrane:2014yza}, they estimated a coupling function with $\kappa=2$, $\beta=2.67$ for LIGO Hanford Observatory (LHO). Measurements made after the second observation run (O2) found $\kappa=0.38$ at LHO and $\kappa=0.25$ at LIGO Livingston Observatory (LLO), and $\beta=3.55, 4.61$~\cite{ligo_logbook} at LHO and LLO respectively. Meanwhile for Virgo, post-O2 measurements indicate $\kappa$=0.275 and $\beta$=2.50~\cite{virgo_logbook}. These measurements highlight that the coupling functions differ in both shape and amplitude at each site.

We made three simplifying assumptions in defining Eq. (\ref{eq:t(f)}), and relaxing each of these assumptions will need to be explored further in future work. For example, it is known that the strength of the coupling function can change as a function of time due to things like routine maintenance on the detectors. Next, recent measurements at LHO indicate that $T(f)$ has a more complicated frequency structure than a simple power law. There is evidence, for example, of a shift to a  positive spectral index near 60~Hz. Finally, the assumption that $T(f)$ is real will also need to be revisited in the future. It could be modeled by multiplying Eq. (\ref{eq:t(f)}) by a frequency-dependent phasor term, $e^{i \phi(f)}$, but there are no measurements at present for the frequency structure of that phase or how it behaves as a function of time. It is possible to generalize the simulations we perform to inject signals that relax these assumptions and evaluate the effect they have on the method we discuss later; however, we reserve such studies for future work.

\subsection{Simulating data}
\label{subsec:simulating_data}
In this section, we first discuss how we generate correlated synthetic magnetometer data streams with a specific overlap reduction function and cross-power. We then discuss how we translate that into strain data using a coupling function. We close with a discussion of the parameters we use to simulate the data.

\subsubsection*{Simulating correlated Gaussian signals}
Here we discuss simulating a correlated Gaussian signal with a specific $M(f)$ and $\gamma^M_{ij}(f)$ between detectors. Let us consider a network of $N$ detectors. Individual on-site magnetometer measurements combine to give an $N$-dimensional column vector, $\bs{\tilde{m}}(f)$, and the magnetic overlap reduction functions are then a hermitian $N \times N$ matrix, $\bs\gamma^M(f)$:
\begin{equation}
\label{eq:general_magnetic_crosscorr}
  \langle \tilde{\bs{m}}(f)\tilde{\bs{m}}^\dagger(f') \rangle \, =  \frac{1}{2}\delta(f-f')\bs\gamma^M(f) M(f).
\end{equation}
The individual elements of the $\bs\gamma^M(f)$ matrix represent the overlap reduction function between different baselines, evaluated at $f$. We then decompose $\bs\gamma^M$ using a Cholesky decomposition~\cite{Cella:2007jh}:
\begin{equation}
    \bs\gamma^M(f) = \bs L(f) \bs L(f)^{\dagger},
\end{equation}
where $\bs L(f)$ is a lower-triangular matrix. We can then use $\bs L(f)$ to construct the correlated magnetometer data,
\begin{equation}
    \tilde{\bs{m}}(f) = \sqrt{\frac{M(f)}{2}} \bs L(f)  \tilde{\bs{\eta}}(f),
\end{equation}
with $\tilde{\bs{\eta}}(f)$ being white Gaussian noise with a covariance matrix given by the identity matrix:
\begin{equation}
     \langle \tilde{\bs{\eta}}(f)\tilde{\bs{\eta}}^{\dagger}(f') \rangle \, =  \textbf{I} \, \delta(f-f').
\end{equation}
Once we obtain $\tilde{\bs{m}}(f)$, which mimic local magnetometer measurements, we project it onto the detectors using a power-law coupling function as in Eq.~(\ref{eq:t(f)}). We then inverse-Fourier transform that strain spectrum to produce $\bs{h}(t)$, and add it to Gaussian detector noise that is uncorrelated between the separate detectors and has a PSD consistent with design sensitivity for the Advanced LIGO and Advanced Virgo detectors~\cite{2018LRR....21....3A}.

\subsubsection*{Correlated magnetic noise PSD and $\gamma^M_{ij}(f)$ for synthetic data sets}

\begin{figure}[h]
    \centering
    \includegraphics[width=.8\linewidth]{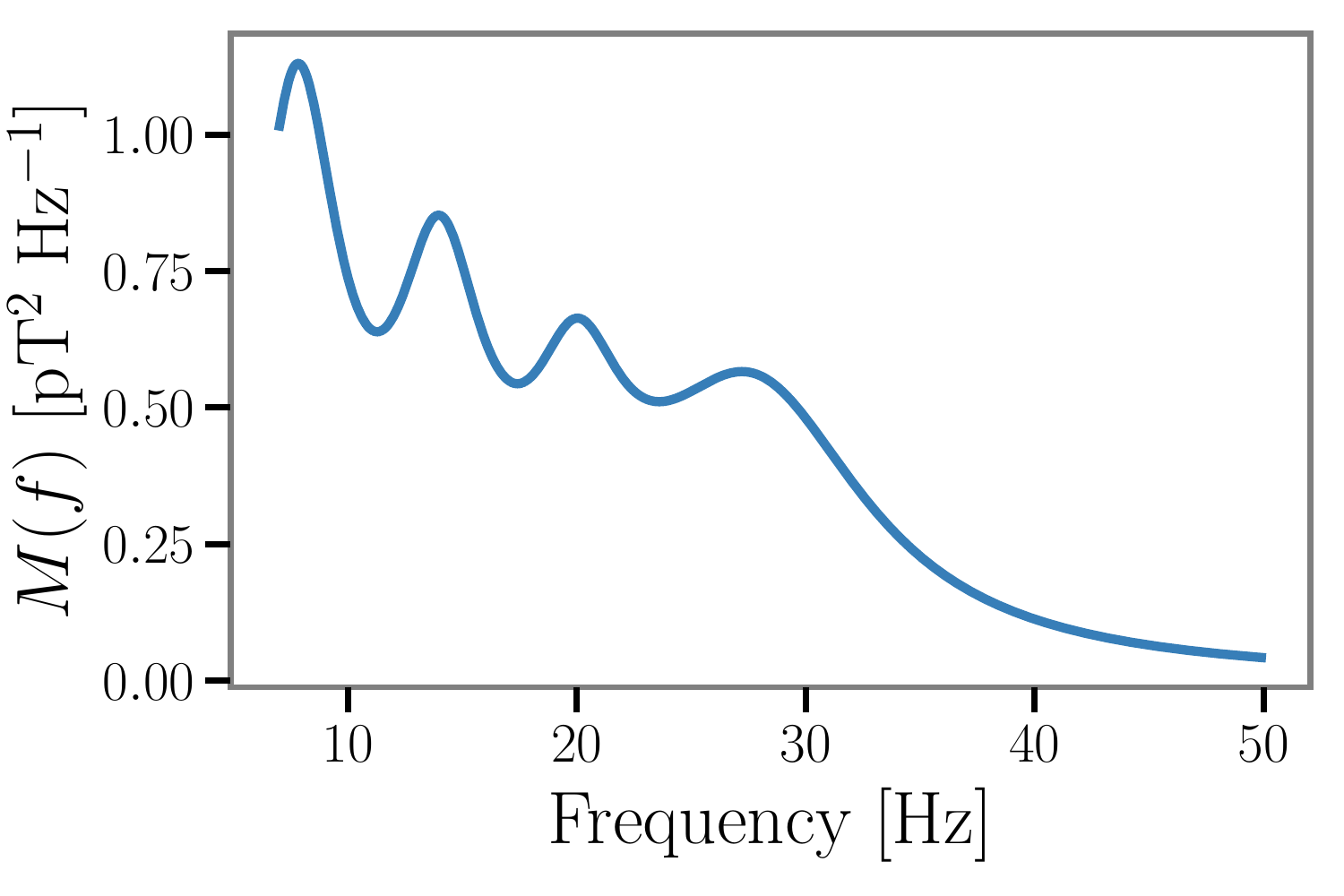}
    \caption{Injected $M(f)$ spectrum. We simulate the first four Schumann peaks as Lorentzians with reasonable amplitudes and widths.}
    \label{fig:m_of_f_plot}
\end{figure}

When constructing a data set with synthetic magnetic noise, we must choose a power-spectral density of the correlated magnetic signal between sites, $M(f)$. This PSD should include the first several harmonics of the Schumann resonances. Throughout the rest of this paper, we model each peak as a separate Lorentzian, with the fundamental peak having an amplitude of $1~\rm{pT^2/Hz}$. A plot of the simulated PSD is shown in Figure~\ref{fig:m_of_f_plot}. We only include harmonics below 30 Hz for this study. While the true correlated magnetic PSD does not fall off as rapidly as our simulated version, the steep coupling functions we consider in Section~\ref{sec:results} will make higher frequencies negligible when the magnetic noise is projected onto the detectors.

{}

\begin{figure}
  \includegraphics[width=0.5\textwidth]{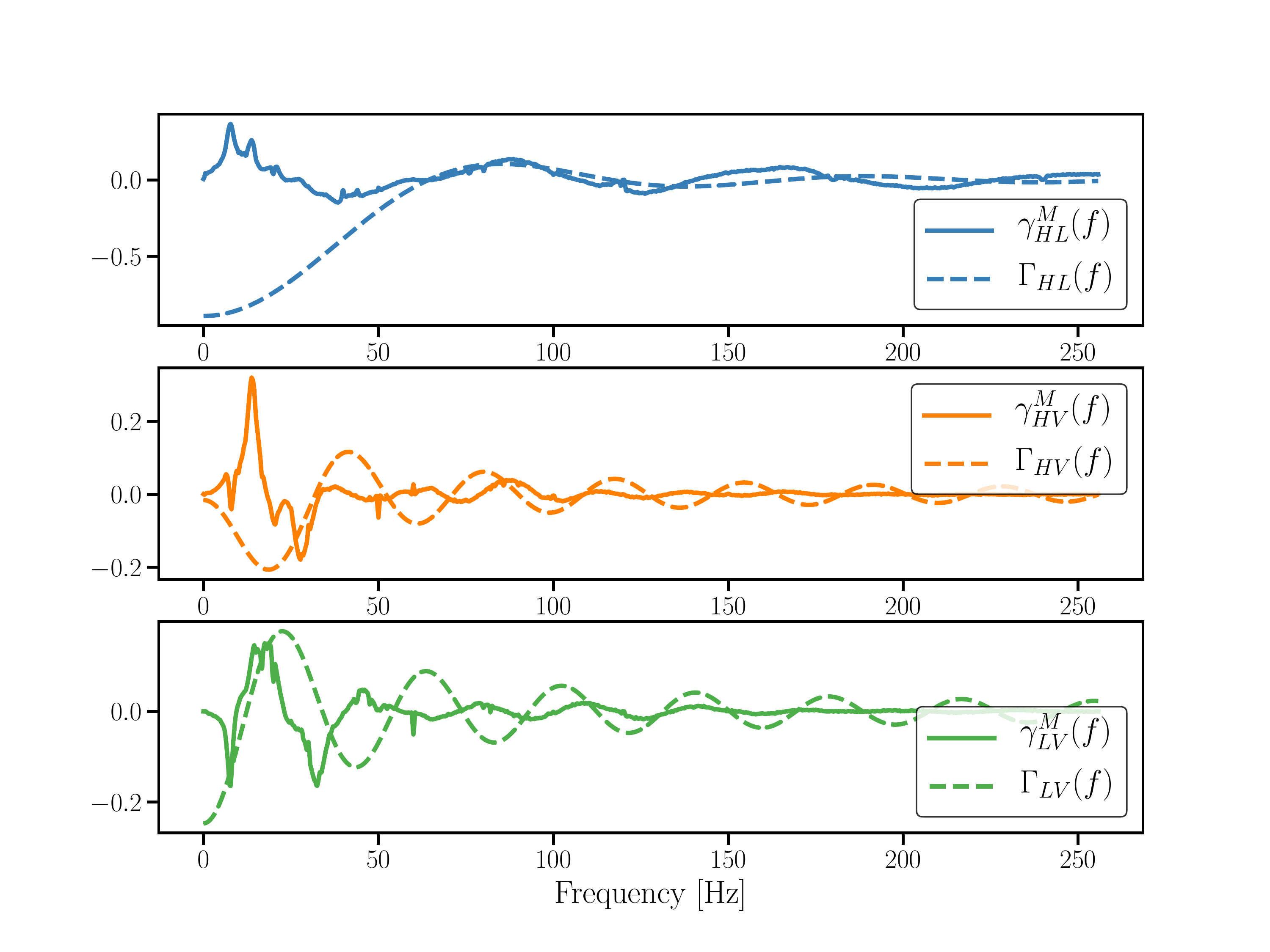}
  \caption{From top to bottom we show $\gamma^M_{ij}(f)$ (solid) and $\Gamma_{ij}(f)$ (dashed) for $ij$=HL, HV, and LV. We discuss how we measure $\gamma^M_{ij}(f)$ in Appendix~\ref{app_A}.}
  \label{fig:overlap_reduction_functions}
\end{figure}

We use the real part of coherence measurements between magnetometers located on site at LHO, LLO and Virgo to estimate $\gamma^M_{ij}(f)$ for each detector pair, and we use these measurements throughout the rest of this paper when creating synthetic data sets. Our use of the real part of the coherences in this case does not affect the results. This can be seen by substituting Eq.~(\ref{eq:transfer}) into Eq.~(\ref{eq:sgwb_searches:c_hat_expectation_value}). A term like Eq.~(\ref{eq:mag_coh}) comes out, multiplied by $T_i(f)$ and $T_j(f)$, which are assumed to be real. A similar, explicit calculation along these lines is done in Section~\ref{sec:pe_ms}. If $T_i(f)$ were not real, then we would need to use the full, complex coherences for $\gamma^M_{ij}(f)$. More details related to these measurements are discussed in Appendix~\ref{app_A}.
A plot of the measured $\gamma^M_{ij}(f)$ is shown for the three detector pairs of interest in Figure~\ref{fig:overlap_reduction_functions}. For comparison, we also include  $\Gamma_{ij}(f)$, which is the analogous quantity for GWs. The differences between $\gamma^M_{ij}(f)$ and $\Gamma_{ij}(f)$ help us to discriminate between correlated magnetic noise and a SGWB in Section~\ref{sec:pe_ms}.

\section{Simultaneous estimation of correlated noise and gravitational waves}
\label{sec:pe_ms}
Various techniques have been proposed to address correlated noise due to Schumann resonances in the output of GW detectors. The most prominent of these techniques is Wiener filtering \cite{Thrane:2013npa,Thrane:2014yza,Coughlin:2016vor,Cirone:2018guh}.
Wiener filtering relies on witness magnetometers that are positioned near the detectors in order to experience the same Schumann resonances, but not too close to be exposed to the local magnetic noise. The downside of Wiener filtering is that it requires a large coherence between the witness and target channels, which means that for weakly coupled signals it can be difficult to completely subtract the noise~\cite{Coughlin:2016vor,Cirone:2018guh}.

We propose an alternative method to address correlated noise, specifically as it pertains to a search for a SGWB. We model the correlated magnetic noise in GW detectors using the data collected by the magnetometers placed near the detector sites, and a parameterized model for the magnetic field to GW detector coupling. We then include this model as a contribution to the estimator, $\hat C_{ij}(f)$ in Eq.~(\ref{eq:sgwb_searches:c_hat_definition}), together with a SGWB model. The way we treat magnetometer data here is reminiscent of the ``a priori'' subtraction scheme presented in~\cite{Thrane:2014yza}, except that here we offer a straightforward way to handle uncertainty in the measurement of the coupling functions by treating them as nuisance parameters that we marginalize over. We reserve a comparison between our method and Wiener filtering for future work.

\subsection{Correlated noise model}

We can rewrite Eq.~(\ref{eq:sgwb_searches:s_gw_definition}) to include separate correlated magnetic and uncorrelated noise terms
\begin{equation}
    \label{eq:s_gw_with_magnetic}
    \tilde s_i(f) = \tilde h_i(f) + \tilde n^{\rm u}_i(f) + T_i(f) \tilde m_i(f),
\end{equation}
where $\tilde n^{\rm u}_i(f)$ is the uncorrelated noise in detector $i$, and $T_i(f) \tilde m_i(f)$ represents the correlated magnetic noise. Substituting Eq.~(\ref{eq:s_gw_with_magnetic}) and Eq.~(\ref{eq:t(f)}) into Eq.~(\ref{eq:sgwb_searches:c_hat_definition}) we find
\begin{equation}
   \label{eq:gw_and_magnetic_contribution_to_point_estimate}
   \langle \hat C_{ij}(f)\rangle = \Omega_{\rm gw}(f) + \Omega_{\textrm M,ij}(f),
\end{equation}
where $\Omega_{\textrm M, ij}(f)$ represents the magnetic contribution, which we derive next.

We construct the magnetic model, $\Omega_{M,ij}(f)$, by first treating local magnetometer data the same way we analyze GW strain data. We break the magnetometer data into $T = 192$~s data chunks, and we calculate the cross-power term in the same way as Eq.~(\ref{eq:sgwb_searches:c_hat_definition}), replacing the strain data with local magnetometer data. That is, for the data between $t_k$ and $t_k +T$ we calculate
\begin{align}
  \hat M_{ij, k}(f) = \frac{2}{T}\frac{\textrm{Re}\left[\tilde m_i^*(f; t_k)\tilde m_j(f; t_k)\right]}{\Gamma_{ij}(f)S_0(f)}.
\end{align}
We  post-process the magnetometer data with the same weights used for post-processing the GW data, viz.
\begin{align}
  \label{eq:m_ij_final_equation}
   \hat M_{ij}(f) &= \frac{\sum_k \hat M_{ij, k}(f) \sigma_{ij,k}^{-2}(f)}{\sum_k \sigma_{ij,k}^{-2}(f)}.
\end{align}
The weights, $\sigma_{ij,k}(f)$, are the same as those expressed in Eq.~(\ref{eq:sgwb_searches:sigma_on_c_hat}). They are calculated using \textit{GW strain data} and not magnetometer data. This way we treat the magnetometer data the same way the magnetic contribution to the final $\hat C_{ij}(f)$ statistic is treated. We then use this final measurement to construct the magnetic contribution to the model, which is given by
\begin{eqnarray}
    \label{eq:y_magnetic}
    \Omega_{\textrm M, ij}(f) &=& \kappa_i\kappa_j\left(\frac{f}{10 \textrm{ Hz}}\right)^{-\beta_i-\beta_j} \hat M_{ij}(f) \times 10^{-22}.
\end{eqnarray}
The factor of $10^{-22}$ assumes that the units of $\tilde m_i(f)$ are $\rm{T~Hz^{-1}}$.

\subsection{Parameter Estimation and Model Selection}
\label{ssec:pe_ms:parameter_estimation}
We use a parameter estimation and model selection scheme similar to those set out in~\cite{Callister:2017ocg,Abbott:2018utx,Mandic:2012iwe}. We choose a Gaussian likelihood for $\hat C_{ij}(f)$ given by
\begin{widetext}
\begin{align}
    \label{eq:pe_ms:likelihood}
    \ln p(\hat C_{ij}(f) | \boldsymbol{\theta}_{\rm gw}, \boldsymbol{\theta}_{\rm M}) = -\frac{1}{2}\sum_{f}\left\{\frac{\left[\hat C_{ij}(f) - \Omega_{\rm gw}(f, \boldsymbol{\theta}_{\rm gw}) - \Omega_{\textrm{M},ij}(f,\boldsymbol{\theta}_{\rm M} )\right]^2}{\sigma_{ij}^2(f)} + \ln\left(2\pi\sigma_{ij}^2(f)\right)\right\},
\end{align}
\end{widetext}
where $\boldsymbol{\theta}_{\rm gw}$ and $\boldsymbol{\theta}_{\rm M}$ represent parameters for the GW and magnetic models respectively. In the case where we have cross-correlation statistics for multiple baselines, we consider the total likelihood to be the product of the individual likelihoods for each pair of detectors. The resulting multi-baseline likelihood is given by
\begin{align}
\label{eq:pe_ms:threedet_likelihood}
    p(\{\hat C_{ij}(f)\}_{ij\in\textrm{pairs}} | \boldsymbol{\theta}_{\rm gw}, \boldsymbol{\theta}_{\rm M}) = \prod_{ij \in \textrm{pairs}} p(\hat C_{ij}(f) | \boldsymbol{\theta}_{\rm gw}, \boldsymbol{\theta}_{\rm M}).
\end{align}

It is straightforward to use Eq.~(\ref{eq:pe_ms:threedet_likelihood}) to estimate the posterior distribution of the parameters, $\boldsymbol{\theta}_{\rm gw}$ and $\boldsymbol{\theta}_{\rm M}$, either by brute-force calculation or by Markov chain Monte Carlo (MCMC) methods~\cite{Gilks1996,PhysRevD.58.082001}.

We will also compare different models for the data using Bayesian model selection. The four models we consider are:
\begin{enumerate}
\item \textbf{NOISE:} $\Omega_{\rm M}(f) = \Omega_{\rm gw}(f) = 0$,
\item \textbf{GW:} $\Omega_{\rm M}(f) = 0$, $\Omega_{\rm gw}(f) \neq 0$,
\item \textbf{SCHU:} $\Omega_{\rm M}(f) \neq 0$, $\Omega_{\rm gw}(f) = 0$,
\item \textbf{GW+SCHU:} $\Omega_{\rm M}(f) \neq 0$, $\Omega_{\rm gw}(f) \neq 0$.
\end{enumerate}
The form of the SGWB model, $\Omega_{\rm gw}(f)$, is the power law in Eq.~(\ref{eq:omega_gw_power_law}), with $\bs{\theta}_{\rm{gw}} =\Omega_{2/3}$ and $\alpha=2/3$ fixed.  The form of $\Omega_{\rm M}(f)$ is given by Eq.~(\ref{eq:y_magnetic}) with $\bs{\theta}_{\rm M}=(\kappa_i, \kappa_j, \beta_i, \beta_j)$ when two detectors are involved. Another set of coupling parameters are included when a third detector is used.

We compare these models using Bayes factors~\cite{skilling2006}. For example, comparing the \textbf{GW} model to the \textbf{NOISE} model we have
\begin{align}
\odds{GW}{NOISE} = \frac{\int \textrm{d}\bs\theta_{\rm gw} p(\hat C_{ij}(f) | \bs\theta_{\rm gw})p(\bs\theta_{\rm gw})}{\mathcal N}\label{eq:bayes_factor_definition}
\end{align}
where $\mathcal N$ is given by evaluating Eq.~(\ref{eq:pe_ms:likelihood}) for $\Omega_{\rm M}(f) = \Omega_{\rm gw}(f) = 0$, and $p(\bs\theta_{\rm gw})$ is the prior on the GW model parameters. When $\odds{GW}{NOISE}>1$ there is support for the \textbf{GW} model compared to the \textbf{NOISE} model. A further discussion of interpretation of Bayes factors can be found in, e.g. chapter 3 of~\cite{Romano:2016dpx}. In this paper, we will consider ``strong'' support for one model over another when $\logodds{}{} > 8$. The numerator of Eq.~(\ref{eq:bayes_factor_definition}) is referred to as the evidence of the \textbf {GW} model and is denoted $\mathcal Z_{\rm GW}$. The prior distribution used for each parameter in the model throughout the rest of this paper is shown in Table~\ref{tab:priors}.

We use the nested sampler~\texttt{CPNest}~\cite{cpnest,skilling2006} through the front-end package \texttt{Bilby}~\cite{Ashton:2018jfp} to both explore the posterior distribution of each parameter and to estimate the evidences for each model.
\begin{table}
  \begin{tabular}{c|c}
    \hline
    \textbf{Parameter} & \textbf{Prior}\\
    \hline
    $\Omega_{2/3}$ & LogUniform($10^{-12}$, $10^{-7}$)\\
    $\kappa_H$ & Uniform(0, 10)\\
    $\kappa_L$ & Uniform(0, 10)\\
    $\kappa_V$ & Uniform(0, 10)\\
    $\beta_H$ & Uniform(0, 10)\\
    $\beta_L$ & Uniform(0, 10)\\
    $\beta_V$ & Uniform(0, 10)\\
    \hline
  \end{tabular}
  \caption{List of prior distributions used for each parameter for results presented in Sections~\ref{sec:results:ms_results}~and~\ref{sec:results:pe_results}.}
  \label{tab:priors}

\end{table}

\section{Results on synthetic data}
\label{sec:results}
In this section we show results for end-to-end simulations of a SGWB search using GW data with correlated magnetic noise. In Section~\ref{ssec:results:simulated_data_parameters} we briefly review  data simulation schemes in the time- and frequency-domains.  In the rest of this section we seek to answer three main questions:
\begin{enumerate}
  \item How does including three detectors aid in our ability to detect the correlated magnetic noise and constrain parameters associated with it?
  \item Can we detect GWs in the context of correlated magnetic noise? How is the significance of the detection affected by the presence of that noise? \label{enum:question1}
  \item Can a noisy measurement of $\hat M_{ij}(f)$ or a strong correlated magnetic signal lead to a false SGWB detection?
\end{enumerate}

\subsection{Synthetic data and parameters}
\label{ssec:results:simulated_data_parameters}
\subsubsection{Time series simulations}
We simulate the strain time-series for the LHO, LLO, and Virgo detectors with correlated magnetic noise using the techniques described in Section~\ref{sec:simulating_data_with_correlated_noise}. We then run the standard pipeline used by LIGO-Virgo for the isotropic search for a SGWB to calculate $\hat C_{ij}(f)$ and $\hat M_{ij}(f)$ for all possible detector pairs.\footnote{\url{https://git.ligo.org/stochastic-public/stochastic}} All SGWB injections are made in the frequency domain on those data products and assume a power law spectrum with $\alpha=2/3$ to mimic an astrophysical SGWB from unresolved CBCs.

The three different year-long synthetic data sets we consider are described in Table~\ref{tab:simulated_data_considered}. We consider data sets with no correlated magnetic noise (\textbf{none}), realistic correlated magnetic noise (\textbf{realistic}) based on post-O2 measurements~\cite{ligo_logbook,virgo_logbook}, and strong correlated magnetic noise (\textbf{strong}). The \textbf{strong} data set corresponds to a larger coupling strength than we currently observe, but is meant to be a stand-in for situations where we do observe correlated magnetic noise. This could occur either due to an increase in the sensitivity of detectors or a change in the coupling functions themselves.

\begin{table}[]
    \centering
    \begin{tabular}{lcccccc}
    \hline
      \textbf{Run name} & $\kappa_{\rm H}$ & $\beta_{\rm H}$ &$\kappa_{\rm L}$ & $\beta_{\rm L}$ & $\kappa_{\rm V}$ & $\beta_{\rm V}$\\
        \hline
    \textbf{None} & 0 & 0 & 0 & 0 & 0 & 0 \\
    \textbf{Realistic} & 0.38 & 3.55 & 0.35 & 4.61 & 0.275 & 2.50 \\
    \textbf{Strong}   & 5 & 3.55 & 5 & 4.61 & 5 & 2.50  \\
        \hline
    \end{tabular}
    \caption{Correlated magnetic noise parameters for four different synthetic data sets.}
    \label{tab:simulated_data_considered}
\end{table}

\subsubsection{Frequency-domain simulations}
\label{sssec:freq_domain_simulations}
For Monte Carlo simulations of many noise realizations we will directly simulate Eq. (\ref{eq:y_magnetic}) in the frequency domain. This simulation method is used in the final two parts of this section, and will also consider the same \textbf{none}, \textbf{realistic}, and \textbf{strong} scenarios detailed in Table~\ref{tab:simulated_data_considered}.

\subsection{Advantages in detecting correlated magnetic noise using three detector network}
We begin by looking at the advantage of having a three-detector, global network as opposed to a simple two-detector network. To evaluate this situation, we use the time-domain data discussed previously. We first look at the effect using three detectors has on model selection, before discussing the advantages of using three detectors when performing parameter estimation.
\subsubsection{Model Selection}
\label{sec:results:ms_results}

In Table~\ref{tab:results:table_no_gw_injection} we show log-Bayes factors comparing different models when there is no injected SGWB. The first column indicates the strength of the correlated noise injection and the second column indicates which detectors were used in the parameter estimation. The other four columns present Bayes factors comparing different models.

The results for the \textbf{none} and \textbf{realistic} injections are shown in the first four rows of Table~\ref{tab:results:table_no_gw_injection}. The log-Bayes factors indicate that there is no preference for a model with correlated magnetic noise compared to Gaussian noise ($\logodds{SCHU}{NOISE}$) or for any model that includes a SGWB compared to Gaussian noise ($\logodds{GW}{NOISE}$ and $\logodds{SCHU+GW}{NOISE}$). Thus, insofar as our simple coupling model is accurate, it is unlikely that at design sensitivity Schumann resonances will be detectable after one year of integration time. However, the coupling functions can change as a function of time, and how they impact the search is highly sensitive to the strength and frequency spectrum of the coupling between the magnetic field and the strain channel of the detector.

The \textbf{strong} injection results are shown in the fifth and sixth rows of Table~\ref{tab:results:table_no_gw_injection}. There is little evidence for correlated magnetic noise with the Hanford-Livingston pair of detectors, but when we include Virgo to the network, we make a clear detection, with $\logodds{SCHU}{NOISE}=33.29$. While we make a detection of Schumann resonances, $\logodds{SCHU+GW}{SCHU}=0.38$ indicates that there is no preference for a model that also includes a SGWB compared to a model that contains just correlated magnetic noise. Including a third detector significantly aids in our ability to detect and characterize correlated magnetic noise in this situation.

\begin{center}
\begin{table*}[ht!]
    \centering
    \begin{tabular}{l|l|c|c|c|c}
    \hline
        \textbf{Run Name} & Dets & $\logodds{GW}{NOISE}$ & $\logodds{SCHU}{NOISE}$ & $\logodds{SCHU+GW}{NOISE}$ & $\logodds{SCHU+GW}{SCHU}$  \\
        \hline
        \textbf{None}       & HL  & -0.65 & -0.26     & -1.0 & -0.74\\
        \textbf{None}       & HLV & -0.75 & 0.45  & -0.32 & -0.77\\
       \textbf{Realistic}   & HL  & -0.61 & -0.36 & -1.01 & -0.65\\
       \textbf{Realistic}   & HLV & -0.57 & -0.53  & -1.18 & -0.65\\
       \textbf{Strong}      & HL  & 0.28 & 0.32  & 0.59  & 0.27\\
       \textbf{Strong}      & HLV & 0.59 & 33.29  & 33.67 & 0.38\\
       \hline
    \end{tabular}
    \caption{We show odds ratios that compare different models when no GW injection is made. We show results for all three injected data sets using just the Hanford (H), Livingston (L) pair, as well as the full Hanford, Livingston, Virgo (V) network.}
    \label{tab:results:table_no_gw_injection}
\end{table*}
\end{center}

\subsubsection{Parameter Estimation}

\label{sec:results:pe_results}
It is also important that we are able to accurately recover the correct parameters for the SGWB, even when there is a strong correlated magnetic noise injection.

\begin{center}
\begin{figure*}[ht!]
    \centering
    \includegraphics[width=0.8\textwidth]{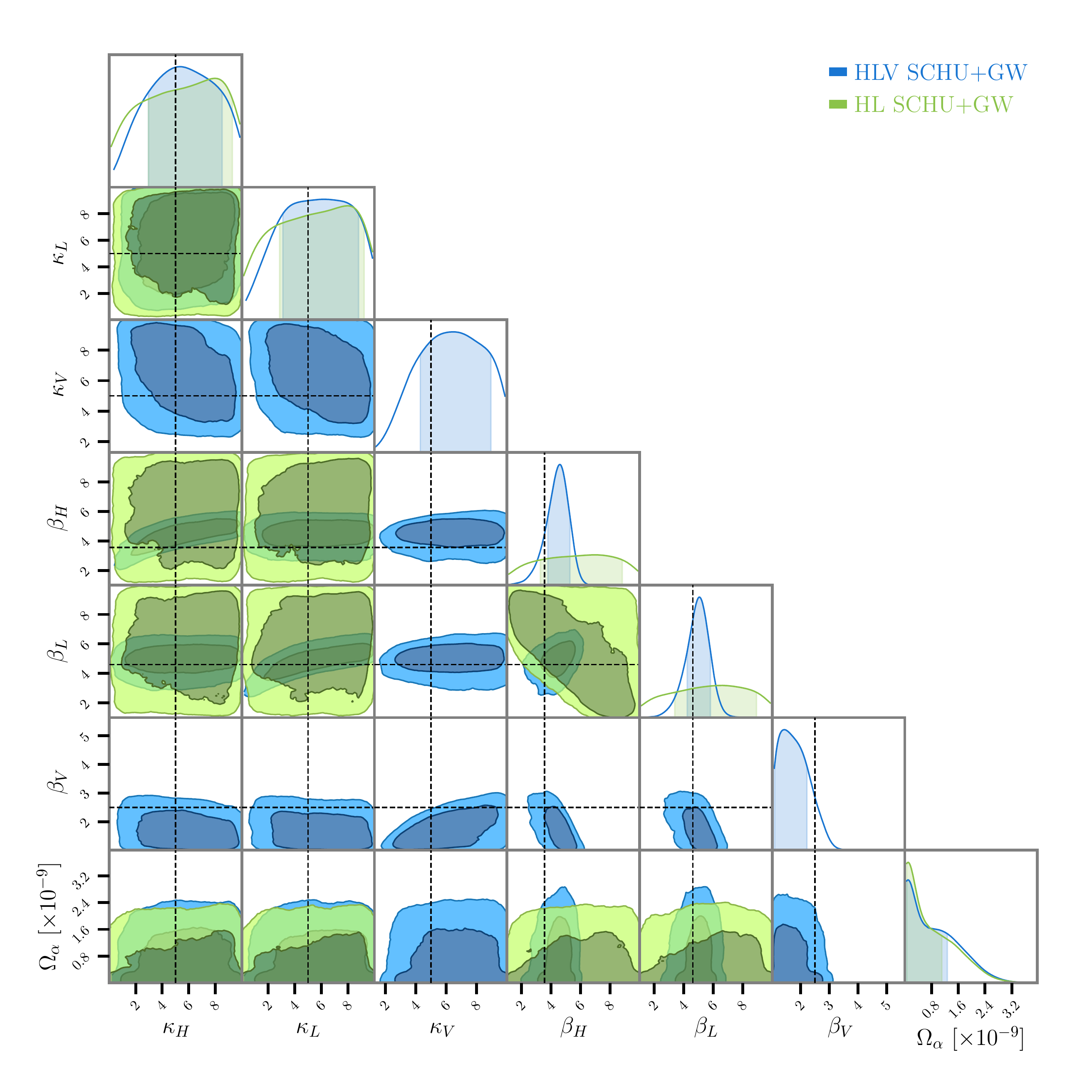}
    \caption{Parameter estimation results for strong correlated noise injection and no SGWB injection. Blue lines and contours correspond to using Hanford, Livingston, and Virgo data. Green lines and contours correspond to using only the Hanford, Livingston pair of detectors. Dashed lines indicate the injected value of each parameter. It is evident that including three detectors improves the recovery of $\kappa$ and $\beta$ for all three detectors. In both cases, the posterior on $\Omega_{2/3}$ is consistent with no SGWB.}
    \label{fig:pe_year_no_injection}
\end{figure*}
\end{center}

In Figure~\ref{fig:pe_year_no_injection}, we show a corner plot with 1- and 2-D marginalized posterior distributions for each parameter over which we sample for the \textbf{strong} injection (last row of Table~\ref{tab:simulated_data_considered}). In this case, there is no SGWB. The green posteriors indicate using only LHO and LLO, while the blue include Virgo in the network as well. It is clear that including Virgo significantly improves our estimates of the Schumann parameters. In the two-detector scenario the magnetic parameters are nearly unconstrained. Whereas, when using the three-detector network, we are able to achieve reasonable estimates of $\beta_H$, $\beta_L$ and $\beta_V$. This makes sense given the model selection results (fifth and sixth rows of Table~\ref{tab:results:table_no_gw_injection}), which indicate that adding Virgo improved our ability to detect correlated magnetic noise. Furthermore, the posterior on $\Omega_{2/3}$ in Figure~\ref{fig:pe_year_no_injection} can be used to set an upper limit on $\Omega_{2/3}$ in the presence of correlated magnetic noise. In Section~\ref{ssec:sgwb_detection}, where we perform frequency domain injections, we will discuss how upper limits on $\Omega_{2/3}$ are affected by the presence of correlated magnetic noise.

\begin{center}
\begin{figure*}[ht!]
    \centering
    \includegraphics[width=0.8\textwidth]{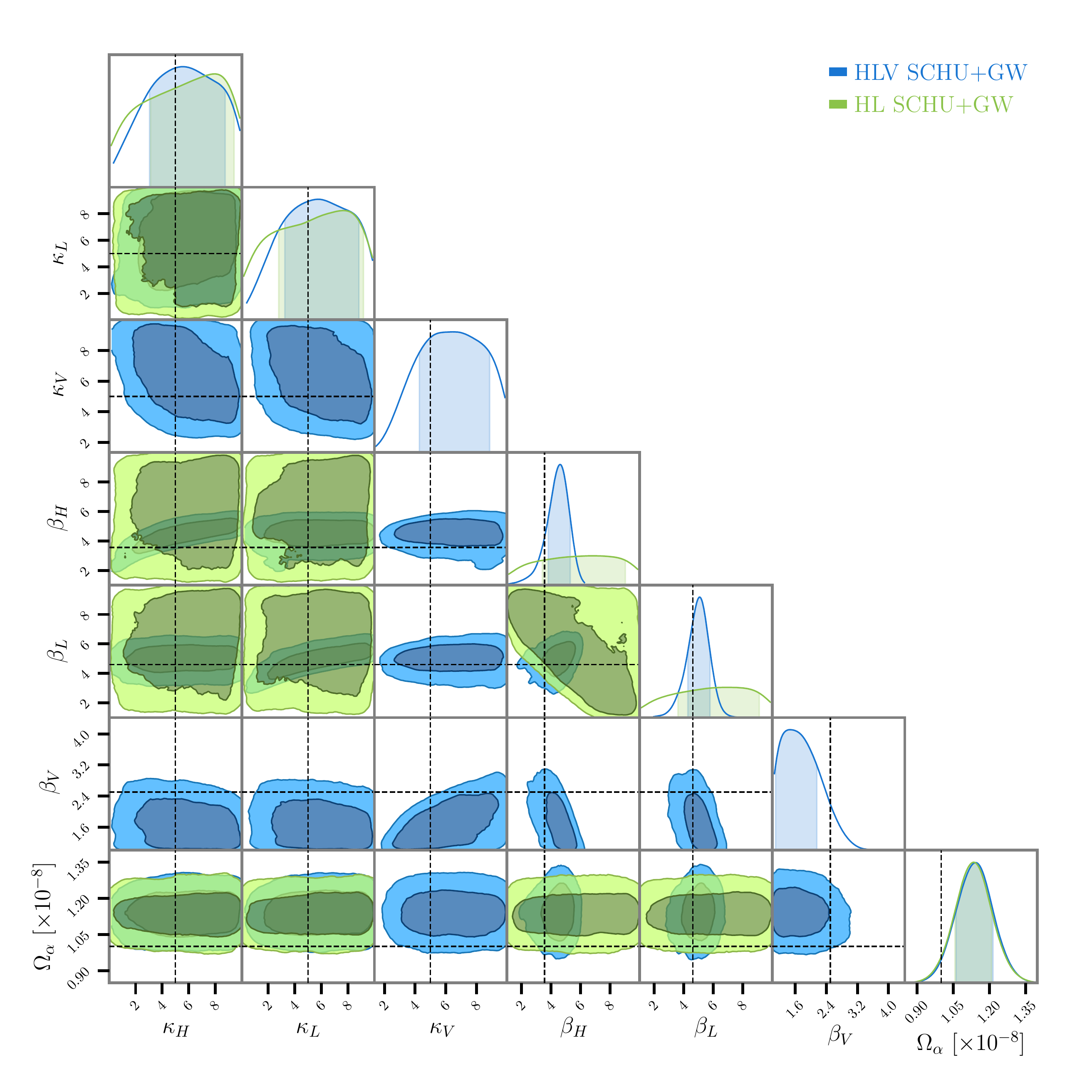}
    \caption{Parameter estimation results for strong correlated noise injection and $\Omega_{2/3} = 10^{-8}$. Blue lines and contours correspond to using Hanford, Livingston, and Virgo data. Green lines and contours correspond to using only the Hanford, Livingston pair of detectors. Dashed lines indicate the injected value of each parameter. It is evident that including three detectors improves the recovery of $\kappa$ and $\beta$ for all three detectors. The posterior distributions of $\Omega_{2/3}$ for both scenarios are consistent with one another, and indicate a 14\% overestimate of $\Omega_{2/3}$.}
    \label{fig:pe_year_with_injection}
\end{figure*}
\end{center}

In Figure~\ref{fig:pe_year_with_injection} we show the same as Figure~\ref{fig:pe_year_no_injection}, but with a SGWB injection of $\Omega_{2/3}=10^{-8}$. The strength of this injection is chosen for illustrative purposes. We see that the posterior on $\Omega_{2/3}$ is well-constrained but represents an overestimate of the true injected value by $14\%$. Due to computational restrictions, we are unable to perform repeated time-domain simulations to evaluate whether this is a systematic bias in our method. However, we did perform repeated frequency-domain simulations (using the method described in Section~\ref{sssec:freq_domain_simulations}) with magnetic and gravitational-wave parameters drawn from the priors in Table~\ref{tab:priors}. Using probability-probability estimates as a diagnostic~\cite{cook:2006ois}, we see no evidence of systematic bias on our estimate of $\Omega_{2/3}$. 

Including Virgo does not improve our ability to constrain $\Omega_{2/3}$. However, it adds significantly to our ability to detect and constrain parameters in the correlated magnetic noise model. A correlated noise detection that is dominated by pairs of detectors that include Virgo is still able to constrain the coupling function parameters in all three detectors, which means that a third detector can aid in our ability to model the correlated noise contribution in the detector pair that is most sensitive to a SGWB.

\subsection{SGWB Detection with correlated magnetic noise}
\label{ssec:sgwb_detection}
\begin{figure}
  \includegraphics[width=0.4\textwidth]{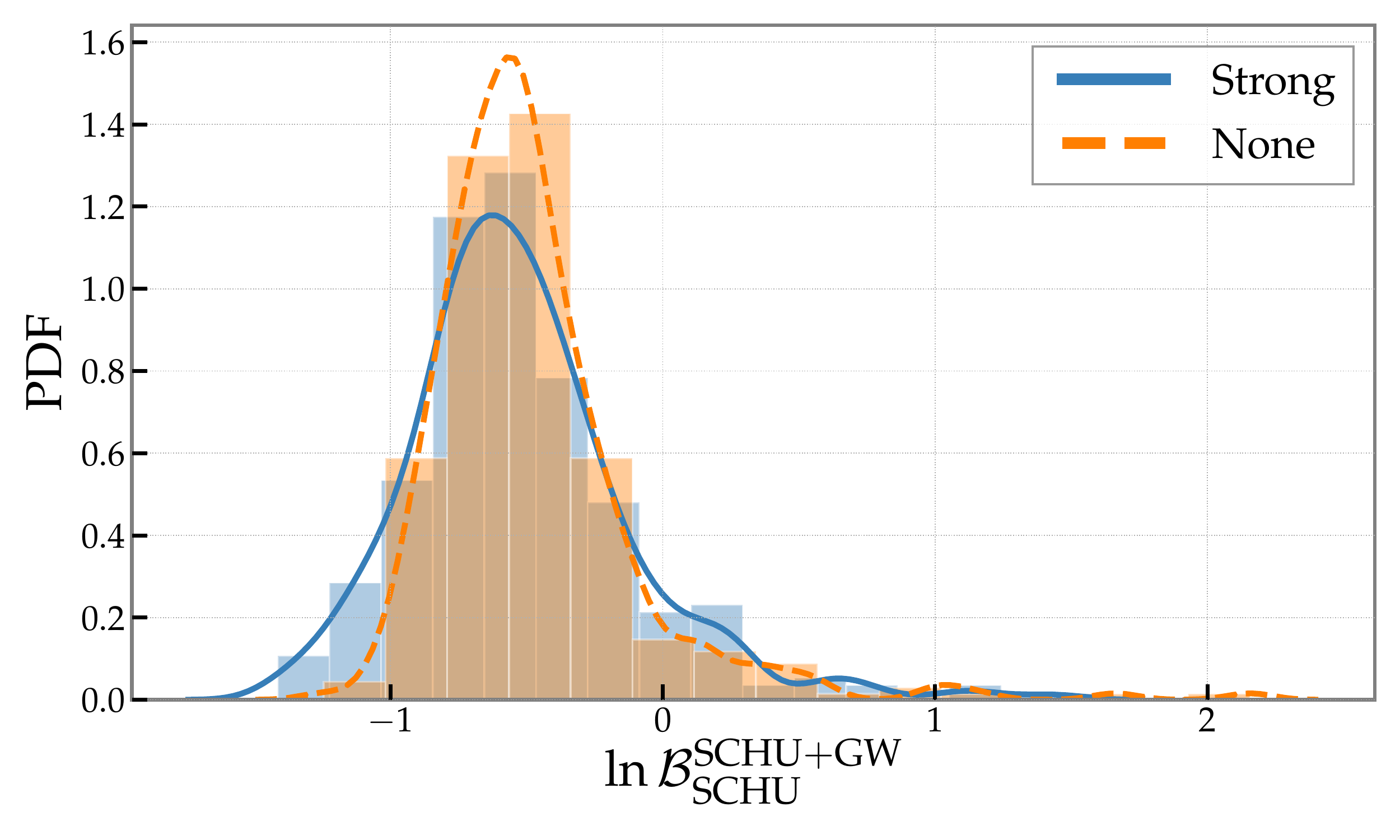}
  \includegraphics[width=0.4\textwidth]{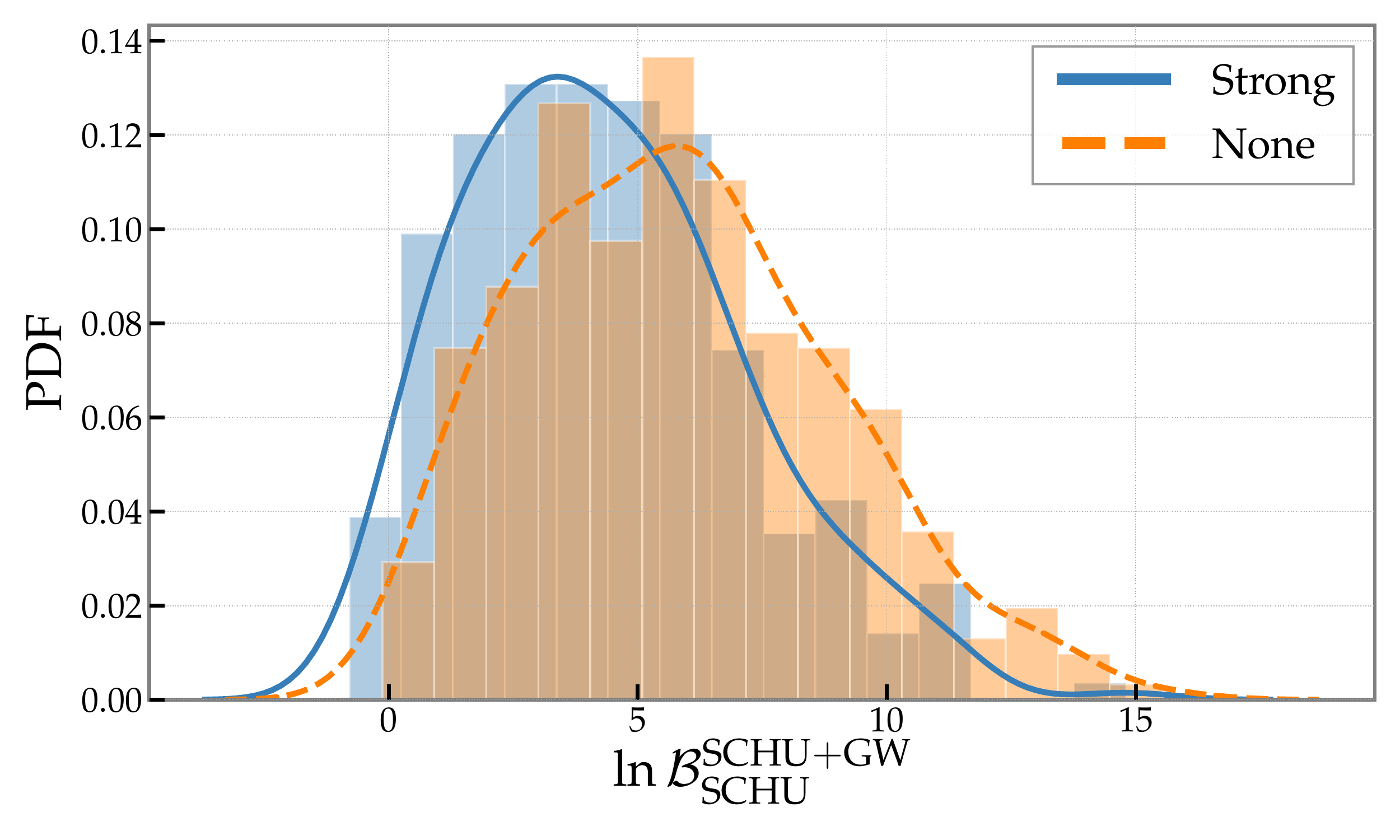}
  \includegraphics[width=0.4\textwidth]{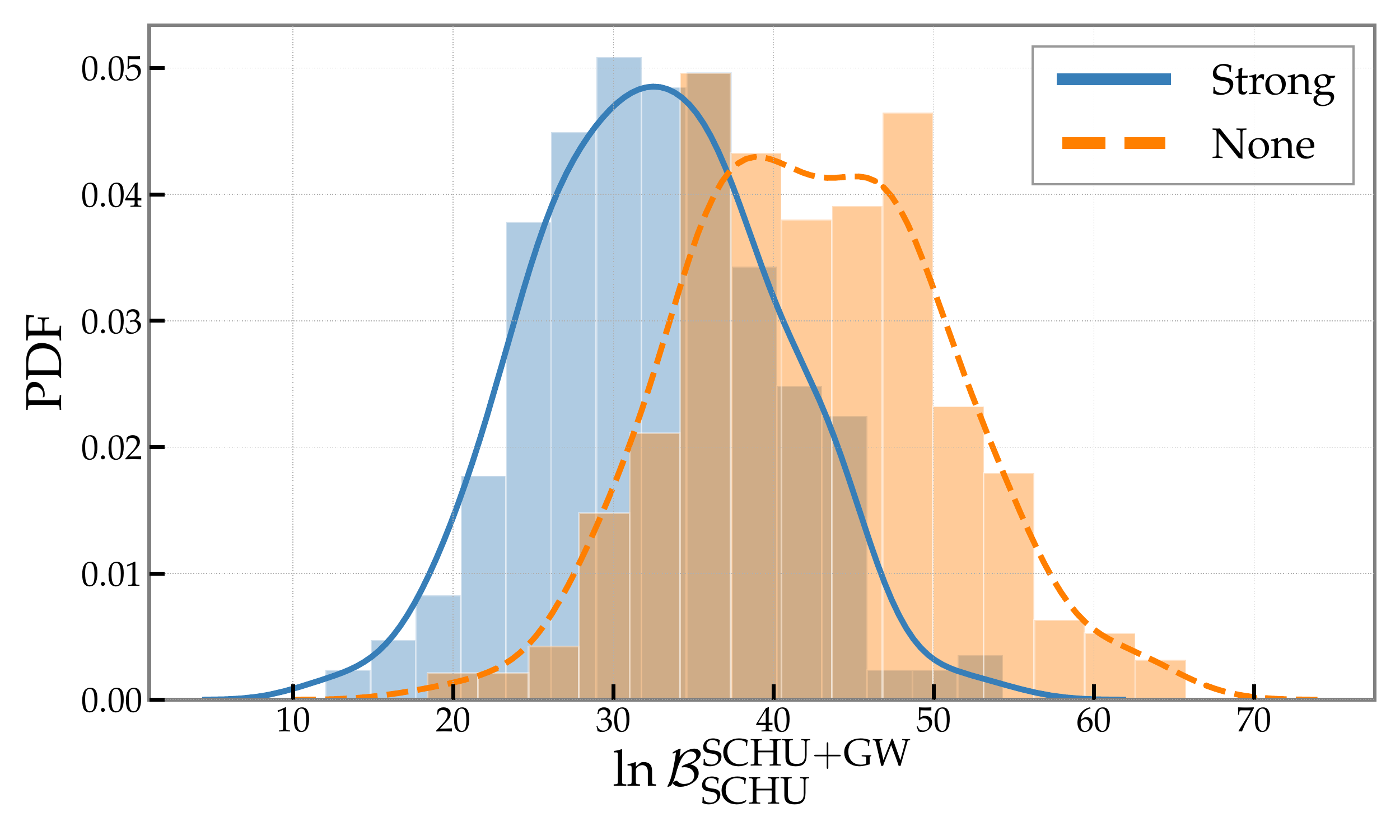}
  \caption{We show the distribution of $\logodds{SCHU+GW}{SCHU}$ for the \textbf{strong} (blue, solid) and \textbf{none} (orange, dashed) injection parameters, and $\Omega_{2/3}=0$ (top), $\Omega_{2/3}=3\times 10^{-9}$ (middle) and $\Omega_{2/3}=10^{-8}$ (bottom). In the top panel we see no evidence for a GW detection. In the middle and bottom panels we see evidence the presence of a SGWB in both cases (although that evidence is marginal in the middle panel). The presence of correlated magnetic noise has clearly shifted the Bayes factor distributions downward.}
  \label{fig:freq_domain_simulation_distributions}
\end{figure}

In this section we show we are able to detect GWs when correlated magnetic noise is present and we show how the presence of correlated magnetic noise affects the significance of that detection.  We performed 300 Monte Carlo simulations, in the frequency domain, of the \textbf{strong} and \textbf{none} correlated noise parameters in Table~\ref{tab:simulated_data_considered}. We did this for $\Omega_{2/3} = 0, 10^{-8}$, and $3\times 10^{-9}$, assuming 1 year of integration time. The results are shown in three panels in Figure~\ref{fig:freq_domain_simulation_distributions}, where we show the distribution of $\logodds{SCHU+GW}{SCHU}$ for each simulation. Throughout this section we use the full three-detector network and all inference is done with the prior distributions in Table~\ref{tab:priors}.

The top panel of Figure~\ref{fig:freq_domain_simulation_distributions}, where $\Omega_{2/3}=0$, shows that for both the strong correlated magnetic noise injection (blue, solid) and the no correlated magnetic noise case we see no preference for the model including GWs compared to the one that only includes correlated magnetic noise, as one would expect. In the absence of a detection of $\Omega_{2/3}$, we can use the posterior distribution on that parameter to set 90\% upper limits for each of the 300 realizations. The median 90\% upper limit on $\Omega_{2/3}$ set for the ensemble of injections is $4.8\times 10^{-10}$ for both the \textbf{strong} and \textbf{none} cases.

In the middle panel of Figure~\ref{fig:freq_domain_simulation_distributions} we show results for $\Omega_{2/3}=3 \times 10^{-9}$, which is within the range of the expected SGWB due to unresolved CBCs~\cite{Abbott:2017xzg}. There is mild evidence for a SGWB for both distributions, with the \textbf{none} distribution (orange, dashed) peaking at $\logodds{SCHU+GW}{SCHU}\approx6$ and the \textbf{strong} distribution (blue, solid) peaking at $\logodds{SCHU+GW}{SCHU}\approx4$. It is clear that when strong correlated noise is present the significance is lower than when there is no correlated noise.

In the bottom panel of Figure~\ref{fig:freq_domain_simulation_distributions} we show results for $\Omega_{2/3}=10^{-8}$. This value is larger than expected for an astrophysical background from unresolved CBCs~\cite{Abbott:2017xzg}, but is chosen for illustrative purposes. When there is strong correlated magnetic noise present (blue, solid) the distribution peaks at a lower value than when there is no correlated magnetic noise injected (orange, dashed), indicating a drop in the significance of the SGWB detection when correlated magnetic noise is present. The median of the simulations with \textbf{strong} correlated magnetic noise is $\logodds{SCHU+GW}{SCHU}=32.2$ compared to $\logodds{SCHU+GW}{SCHU}=42.2$ for the \textbf{none} simulation, corresponding to a 31\% drop in the detection statistic.

\begin{figure}
    \centering
    \includegraphics[width=0.49\textwidth]{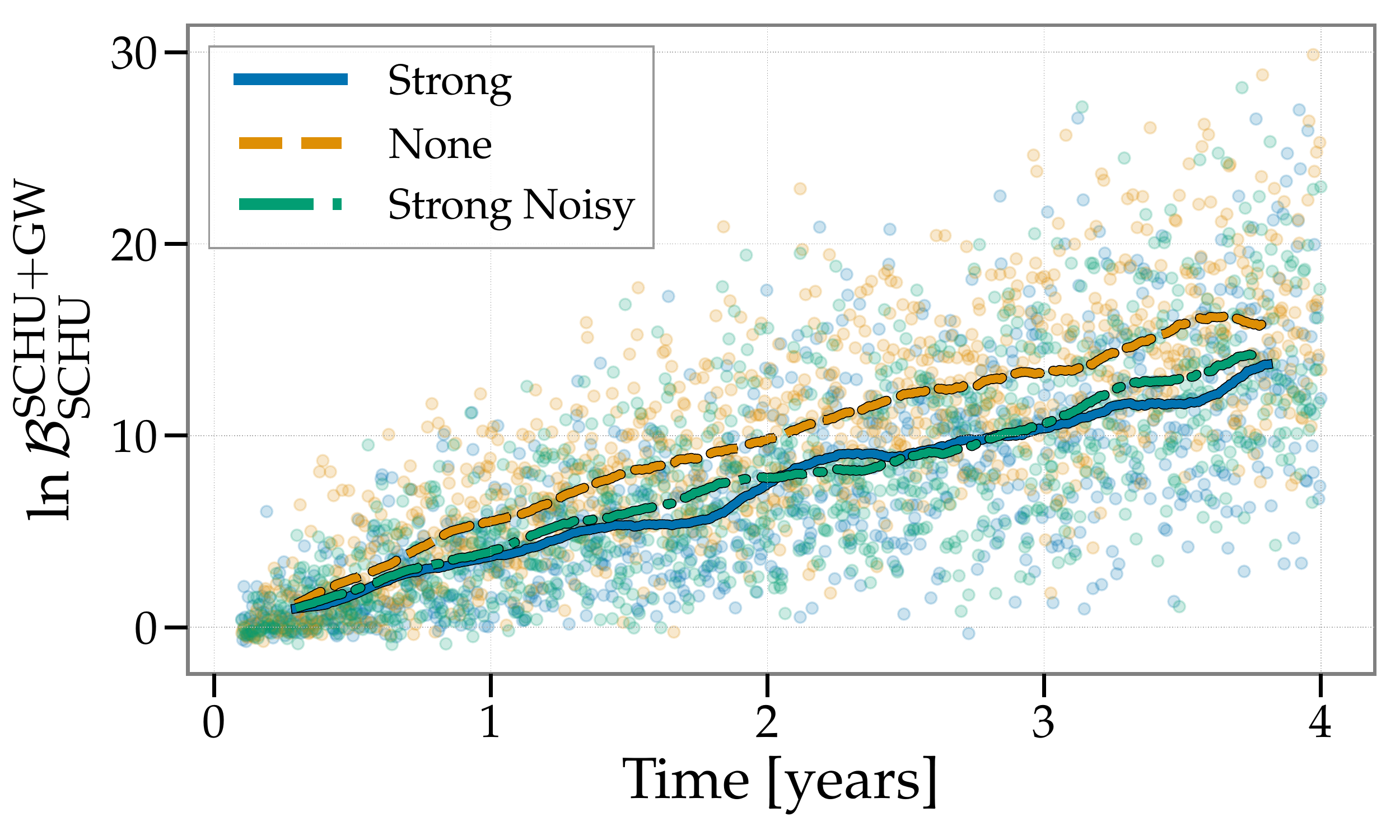}
    \caption{We show how $\logodds{SCHU+GW}{SCHU}$ scales as a function of time using 1000 injections in the frequency domain with increasing observation time with a SGWB injection of $\Omega_{2/3}=3\times 10^{-9}$. The \textbf{strong} case (blue, solid) is clearly below the \textbf{none} case (orange, dashed). We also show the \textbf{strong} case with a noisy measurement of $\hat M_{ij}(f)$ with a magnetic SNR of 5 (green, dash-dot). It does not appear that a noisy measurement of $\hat M_{ij}(f)$ significantly hinders our ability to detect a SGWB.}
    \label{fig:time_dependent_bayes_factors}
\end{figure}

Figure~\ref{fig:freq_domain_simulation_distributions} shows that the presence of correlated magnetic noise reduces the significance of a GW detection. In Figure~\ref{fig:time_dependent_bayes_factors} we show how $\logodds{SCHU+GW}{SCHU}$ scales with time for the \textbf{strong} (blue, solid) and \textbf{none} (orange dashed) cases with an injection of $\Omega_{2/3}=3\times  10^{-9}$.  We also show a third case where we consider a noisy measurement of $\hat M_{ij}(f)$, which we will discuss in Section~\ref{ssec:noisy_measurement}. The \textbf{strong} and the \textbf{none} cases are clearly different, and the time-to-detection (in this case the time to reach $\logodds{SCHU+GW}{SCHU}=8$) is increased to \TimeToDetectionStrong~for the \textbf{strong} case compared to \TimeToDetectionNoSchumann~for the \textbf{none} case (values given define the 68\% confidence regions).

\subsection{Can a poor measurement of $\hat M_{ij}(f)$ lead to a false SGWB detection?}
\label{ssec:noisy_measurement}
To this point, we have not considered the effect of local magnetometer noise, which can reduce the significance with which we measure the noise that is correlated between the detectors. In this section, we address whether a low-SNR measurement of $\hat M_{ij}(f)$, defined in Eq.(\ref{eq:m_ij_final_equation}), or very strong correlated noise could lead to a false GW detection. To evaluate this question, we perform frequency-domain injections with increasing values of $\kappa$ from  0 to 9 for each detector, and the same $\beta$ values for LHO, LLO, and Virgo that were used for the \textbf{realistic} and \textbf{strong} injections in Table~\ref{tab:simulated_data_considered}. For this test, we use the full three detector network and we extend the upper range of the priors, shown in Table~\ref{tab:priors}, on $\kappa$ from 10 to 20.

We also vary the confidence with which we measure $\hat M_{ij}(f)$. We perform our frequency domain injection using $M(f)$ presented in Figure~\ref{fig:m_of_f_plot}. We then simulate a ``measurement'' at a chosen signal-to-noise (SNR) in each frequency bin by drawing $\hat M_{ij}(f)$ from a normal distribution with mean $\gamma^{M}_{ij}(f) M(f)$ and variance $\sigma^2_M(f)$. Due to the fact that the SNR and $M(f)$ are chosen a priori, we re-arrange the definition of the SNR to set the standard deviation in each frequency bin,
\begin{align}
    \label{eq:magnetic_sigma}
  \sigma_M(f) &= \frac{\gamma_{ij}^M(f) M(f)}{\textrm{SNR}}.
\end{align}
We perform frequency domain injections with SNRs ranging from 1 to 35.

\begin{figure}
  \includegraphics[width=0.5\textwidth]{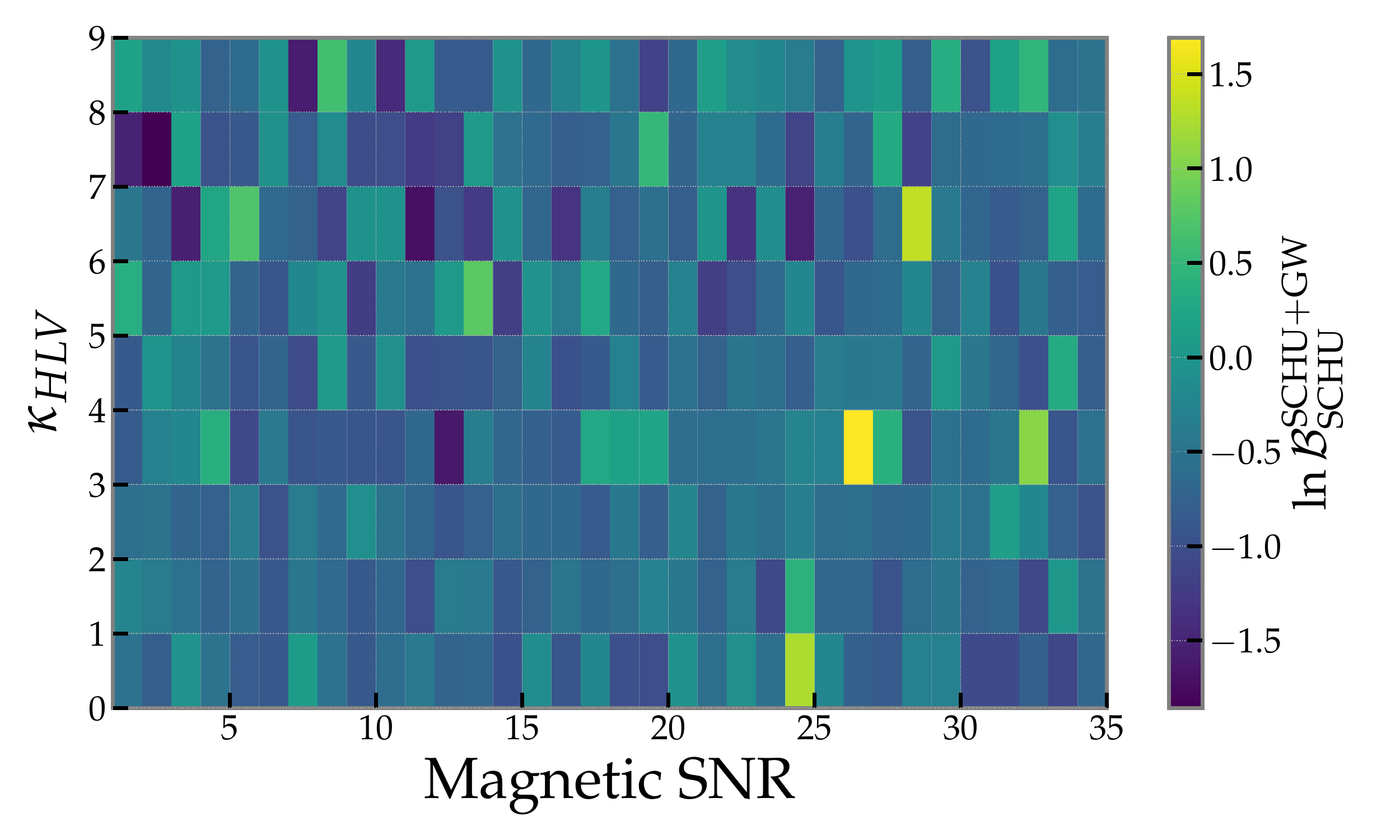}
  \caption{We show a grid of $\logodds{SCHU+GW}{SCHU}$ for different values of $\kappa$ (the same $\kappa$ is used for all three detectors) and the magnetic SNR defined in Eq. (\ref{eq:magnetic_sigma}). The range of $\logodds{SCHU+GW}{SCHU}$ across the whole grid is consistent with no SGWB detection. This indicates it is unlikely that a false SGWB detection could be caused by a noisy measurement of $\hat M_{ij}(f)$ or the presence of strong correlated magnetic noise.}
  \label{fig:parameter_space_test}
\end{figure}

In Figure~\ref{fig:parameter_space_test} we show $\logodds{SCHU+GW}{SCHU}$ for the range of $\kappa$ and magnetic SNR values we inject and with $\Omega_{2/3}=0$. The Bayes factors in Figure~\ref{fig:parameter_space_test} are consistent with no detection--they span a similar range to those in the top panel of Figure~\ref{fig:freq_domain_simulation_distributions}, where we assumed a perfect measurement of $\hat M_{ij}(f)$. This result indicates that a false detection of a SGWB is unlikely, even with an uncertain measurement of the Schumann resonances.

We also test whether a noisy measurement of $\hat M_{ij}(f)$ could increase time-to-detection of a SGWB. We do this by showing how $\logodds{SCHU+GW}{SCHU}$ scales with time for $\kappa=5$ (\textbf{strong} case) and SNR=5 in Figure~\ref{fig:time_dependent_bayes_factors} with the green dash-dot curve. There is a not a clear reduction in detection strength compared to the blue solid curve, which is the same correlated magnetic noise strength but with a no-noise measurement of $\hat M_{ij}(f)$. The time-to-detection for the noisy measurement case is~\TimeToDetectionStrongNoisy, which is around 50\% longer than the case with no correlated noise and comparable to the case where we make a noiseless measurement of the magnetic field.

\section{Discussion}
\label{sec:discussion}
In this paper we perform realistic realistic simulations of correlated magnetic noise in interferometric gravitational-wave detectors, and propose a new method to detect a SGWB in the presence of that correlated magnetic noise. The method reliably separates a SGWB from correlated magnetic noise, although significance of that detection can be reduced either by the presence of strong correlated noise, or through a noisy measurement of the correlated magnetic fields. We also showed that a three-detector network improves our ability to detect, estimate, and subtract the correlated magnetic noise compared to just a single detector pair. Moreover, in the absence of a SGWB detection, upper limits on the SGWB are a natural byproduct of the analysis.

The method presented here is an alternative to Wiener filtering, but could also be used in tandem with Wiener filtering. For example, this method could be used to find correlated noise not successfully subtracted using Wiener filtering. A full comparison of the efficiency of this method compared to Wiener filtering is reserved for future work. Moreover, any proposed SGWB signal could be verified using the geodesy methods discussed in Ref.~\cite{Callister:2018ogx}. In that scenario, the maximum a posteriori parameters could be used to subtract off the correlated magnetic noise, and the proposed remaining SGWB signal could be analyzed using geodesy.

This method is easily applicable to current searches for a SGWB, and should help make a reliable detection of a SGWB using ground-based interferometric detectors. Future work should focus on using a model for the magnetic coupling functions that is more flexible than a simple power law, making direct comparisons with other proposed methods, and working towards incorporating the time-variability of both coupling functions and Schumann resonances.

\section{Acknowledgements}
The authors would like to thank Thomas Callister, Giancarlo Cella, and the LIGO/Virgo Stochastic Background group for helpful comments and discussions. The authors would also like to thank the scientists on site at the Virgo and LIGO detectors for installation and maintenance of the low noise magnetometers whose data we used in this paper.
 Parts of this research were conducted by the Australian Research Council Centre of Excellence for Gravitational Wave Discovery (OzGrav), through project number CE170100004.
 K.M. is supported by King's College London through a Postgraduate International Scholarship.
 N.C. acknowledges support from National Science Foundation grant PHY-1806990. M.S. is supported in part by the Science and Technology Facility Council (STFC), United Kingdom, under the research grant ST/P000258/1. This paper has been given LIGO DCC number P2000258.

 Numerous software packages were used in this paper. These include \texttt{matplotlib}~\cite{Hunter:2007}, \texttt{numpy}~\cite{numpy}, \texttt{scipy}~\cite{2020SciPy-NMeth}, \texttt{bilby}~\cite{Ashton:2018jfp}, \texttt{cpnest}~\cite{cpnest}, \texttt{ChainConsumer}~\cite{Hinton2016}, \texttt{seaborn}~\cite{michael_waskom_2014_12710}.

\bibliography{your bibliography}
%
\appendix

\section{Simulated magnetic noise properties}
\label{app_A}
\begin{figure}[t!]
        \centering
        \includegraphics[width=0.3\textwidth]{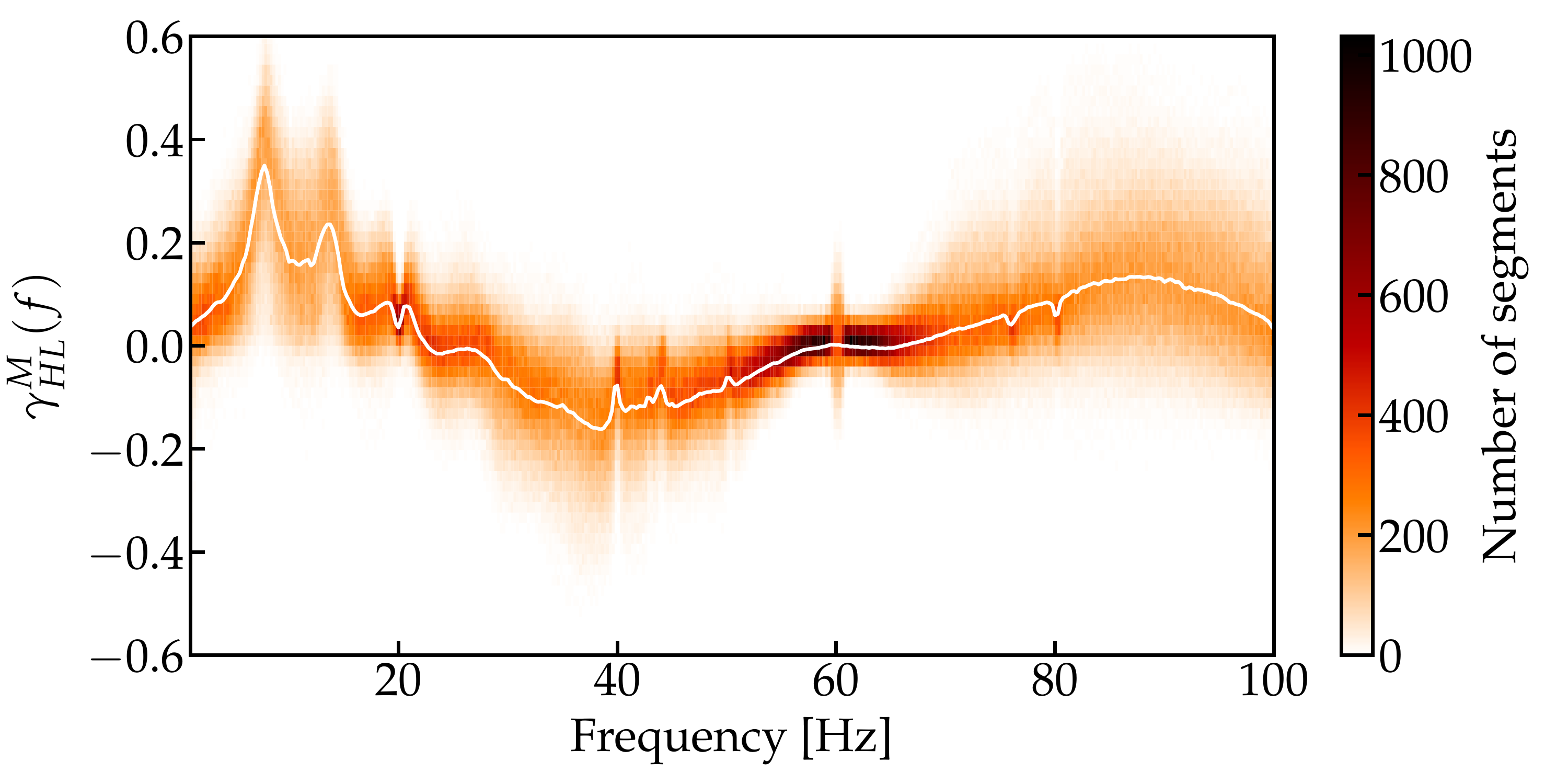}
        \includegraphics[width=0.3\textwidth]{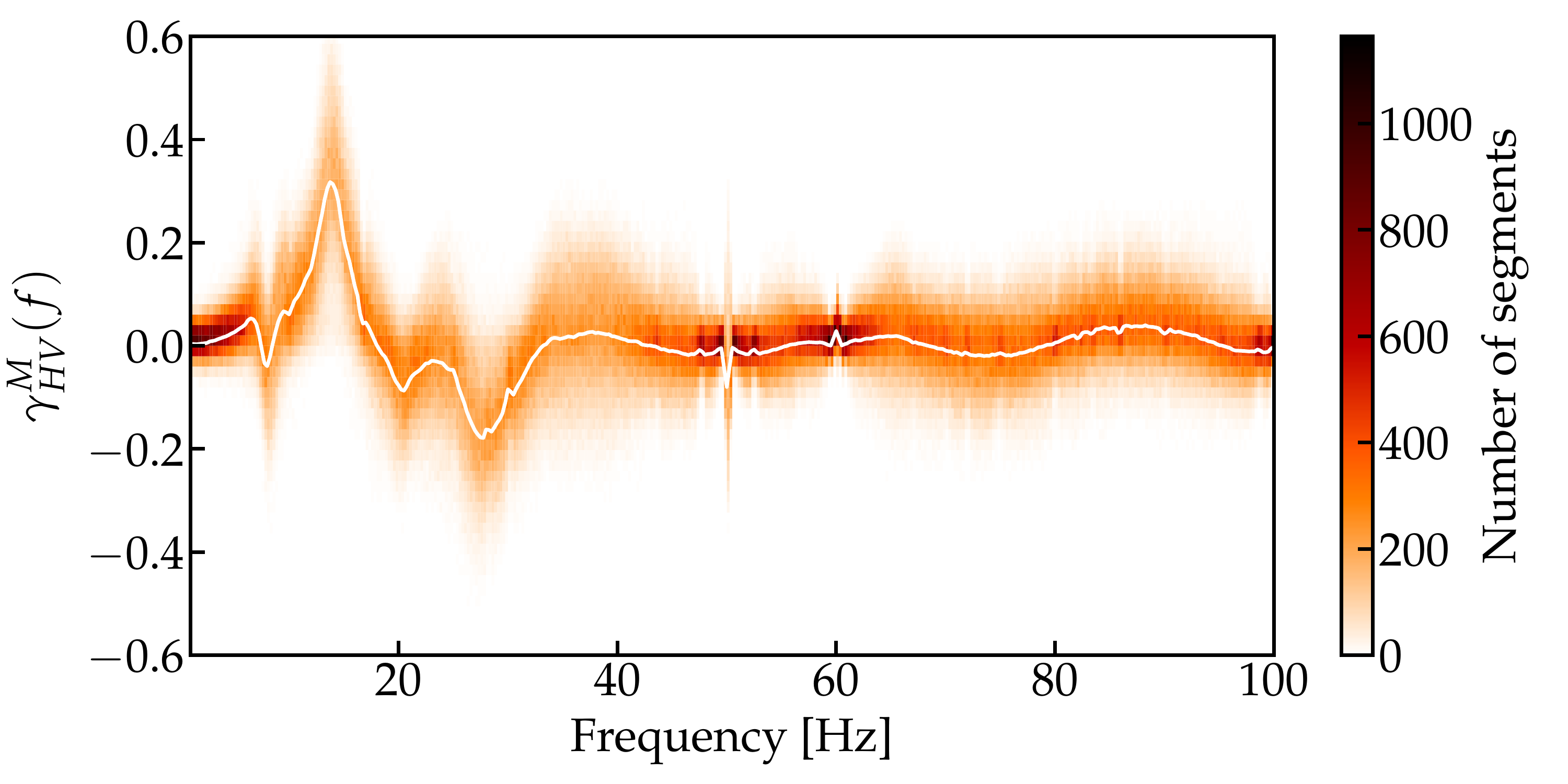}
        \includegraphics[width=0.3\textwidth]{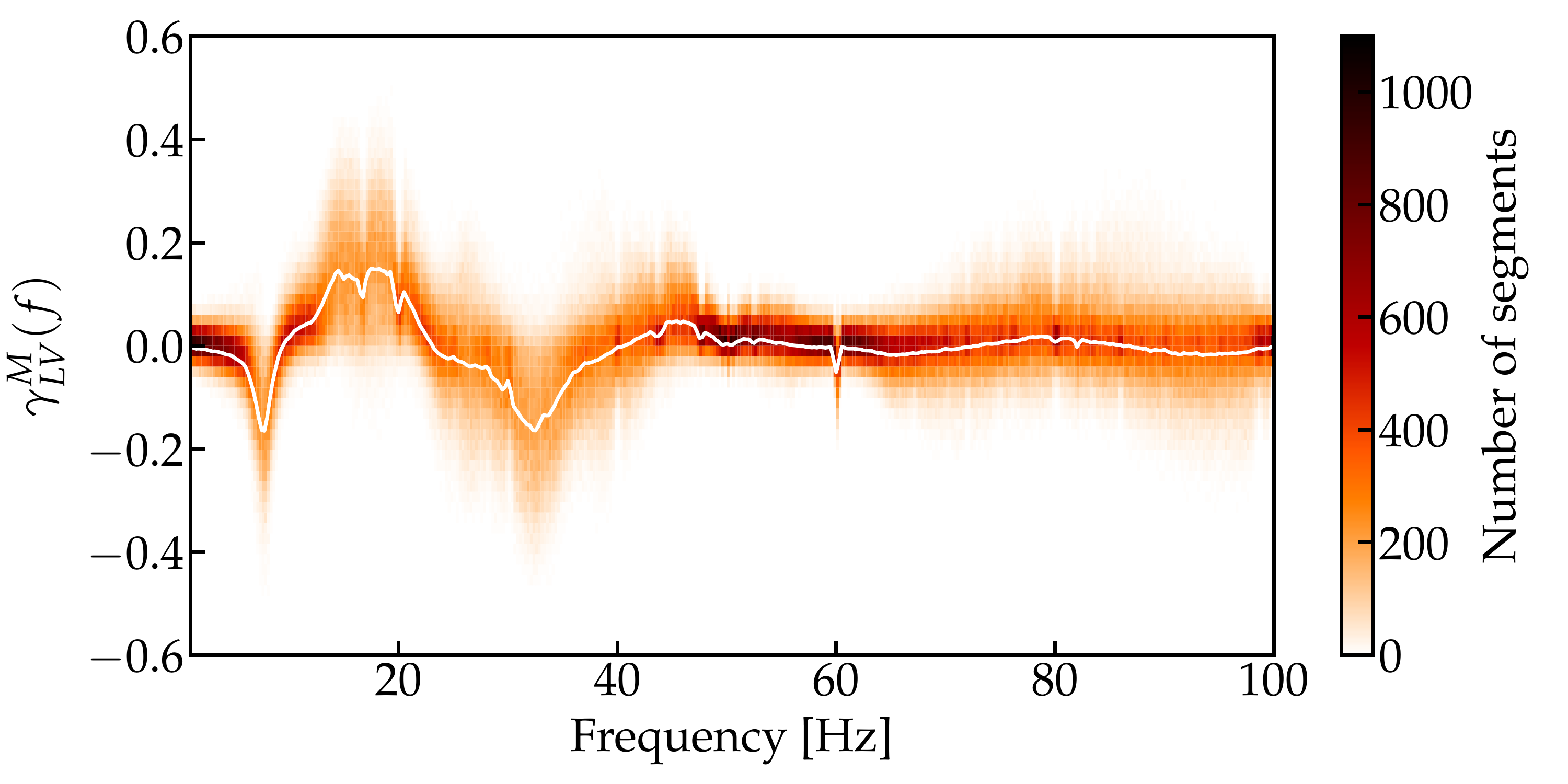}
        \caption{Color indicates histogram of RPCC of Hanford-Livingston for each 1800~s chunk of data available for 60~days. The median value of all RPCC measurements taken at each frequency is shown in white.}
        \label{fig:hl_rppc_orf}
    \end{figure}

We use low noise magnetometers on-site at the Advanced LIGO and Advanced Virgo detectors and correlate them to deduce what $\gamma^M_{ij}$, defined in Eq.~(\ref{eq:mag_coh}), looks like. A discussion of the magnetometers and their locations is given in~\cite{Cirone:2018guh}. We use the real part of complex coherence (RPCC), defined as
\begin{align}
\gamma^M_{ij}(f;t) = \textrm{Re}\left[\frac{\tilde m_i^*(f;t) \tilde m_j(f;t)}{\sqrt{\tilde m_i^*(f;t)\tilde m_i(f;t)}\sqrt{\tilde m_j^*(f;t)\tilde m_j(f;t)}}\right]
\end{align}
where $\tilde m_i(f;t)$ is the Fourier transform of the data from magnetometer $i$ starting at time $t$ evaluated at frequency $f$.
We calculate the numerator and denominator of $\gamma^M_{ij}(f;t)$ separately over 4~s segments and average them separately over 1800~s of data to create an estimate of $\gamma^{M}_{ij}(f;t)$ for that 1800~s chunk of data. We do this for each 1800~s chunk of data available from from~July 9, 2019 00:00 UTC -- September 7 2019 00:00 UTC.
We then take a histogram at each frequency over all of the 1800~s measurements. A heatmap of this histogram is shown in Figure~\ref{fig:hl_rppc_orf} for each possible detector pair.
For the simulations discussed in~Section~\ref{sec:results}, we use the median over the time chunks at each frequency, indicated by the white line in each panel in Figure~\ref{fig:hl_rppc_orf}. This is indicated by the white line in Figure~\ref{fig:hl_rppc_orf}.

The RPCC is not an exact measurement of $\gamma^{\rm M}_{ij}(f)$. It approximates this value only insofar as the the ``signal'', $M(f)$, dominates the noise in the individual magnetometers. However, in the absence of a reliable analytic calculation (which is available in the GW case, for example), it is a good heuristic for capturing the sign and general shape of $\gamma^{\rm M}_{ij}(f)$.

\end{document}